\documentclass[journal,10pt,twocolumn,twoside]{IEEEtran}
 
\usepackage[dvips]{graphicx}
\usepackage[utf8]{inputenc}
\usepackage[T1]{fontenc}
\usepackage{diagbox}

\usepackage{amsmath,amssymb,amsfonts,amsthm,mathrsfs}
\usepackage{epsfig,epsf,subfigure,graphicx,graphics}
\usepackage{url,enumerate}
\usepackage{color}
\usepackage{cite}
\usepackage{dblfloatfix,fixltx2e}
\usepackage{CJK,tikz}
\usepackage{multirow}
\usepackage{algorithm,algorithmic}
\begin{document}

\title{Circularly Pulse-Shaped Precoding for OFDM:\\ A New Waveform and Its Optimization Design for 5G New Radio}

\author{Yenming~Huang,~\IEEEmembership{Student~Member,~IEEE,}
    and~Borching~Su,~\IEEEmembership{Member,~IEEE}
\thanks{Y. Huang and B. Su are 3GPP standardization delegates of National Taiwan University, and are with Graduate Institute of Communication Engineering, National Taiwan University, Taipei 10617, Taiwan (e-mail: d01942015@ntu.edu.tw; borching@ntu.edu.tw).}
    }


\maketitle

\begin{abstract}
A new circularly pulse-shaped (CPS) precoding orthogonal frequency division multiplexing (OFDM) waveform, or CPS-OFDM for short, is proposed in this paper.
CPS-OFDM, characterized by user-specific precoder flexibility, possesses the advantages of both low out-of-subband emission (OSBE) and low peak-to-average power ratio (PAPR), which are two major desired physical layer signal properties for various scenarios in 5G New Radio (NR), including fragmented spectrum access, new types of user equipments (UEs), and communications at high carrier frequencies. 
As opposed to most of existing waveform candidates using windowing or filtering techniques, CPS-OFDM prevents block extension that causes extra inter-block interference (IBI) and envelope fluctuation unfriendly to signal reception and power amplifier (PA) efficiency, respectively. 
An optimization problem of the prototype shaping vector built in the CPS precoder is formulated to minimize the variance of instantaneous power (VIP) with controllable OSBE power (OSBEP) and noise enhancement penalty (NEP). 
In order to solve the optimization problem involving a quartic objective function, the majorization-minimization (MM) algorithmic framework is exploited.
By proving the convexity of the proposed problem, the globally optimal solution invariant of coming data is guaranteed to be attained via numbers of iterations. 
Simulation results demonstrate the advantages of the proposed scheme in terms of detection reliability and spectral efficiency for practical 5G cases such as asynchronous transmissions and mixed numerologies.  
\end{abstract}



\section{Introduction}
\label{Sec:Introduction}
The Third Generation Partnership Project (3GPP) has initiated the standardization activity for the fifth generation (5G), officially named as New Radio (NR), since 2016. The new-air interface has been envisioned to support diverse use cases, broadly classified as enhanced mobile broadband (eMBB), ultra-reliable low-latency communications (URLLC), and massive machine type communications (mMTC), operated in a wide range of frequencies and deployment scenarios \cite{3GPPTR38913}. Waveforms of NR, on basis of orthogonal frequency division multiplexing (OFDM) \cite{3GPPTR38912}, are being developed to flexibly address various emerging applications and physical layer signal requirements by providing different transmission properties. Support of discrete Fourier transform spread OFDM (DFT-S-OFDM) based waveforms is mandatory for user equipments (UEs) in view of link budget \cite{3GPPTR38912}. 

OFDM has achieved great success in Long Term Evolution (LTE) of the fourth generation (4G) due to its several merits such as robustness to channel frequency selectivity, plain channel estimation, flexibility in frequency domain multiple access, easy integration with multiple-input multiple-output (MIMO) technologies, etc \cite{Sesia2011_TB}. 
However, different from 4G LTE, there are new application scenarios such as fragmented spectrum usage including asynchronous transmissions and mixed numerologies, low-cost machines, and communications at high carrier frequencies in upcoming 5G wireless systems \cite{Wunder2014_WD, Guan2017_WD, Zaidi2016VTC_WD}. 
Two additional demands, namely, spectral containment and resistance to power amplifier (PA) nonlinearity, emerge to fit these new scenarios\cite{Lien2017_HYM, Zaidi2016_WD}.
The first is to lower the out-of-subband emission (OSBE) of a user assigned to a portion of OFDM subcarriers in a carrier. In this way, the incurring sidelobe leakage interference imposed on frequency domain adjacent users in the carrier, for lack of orthogonality, can be mitigated. It makes relaxing stringent synchronization requirements as specified in 4G LTE possible, so as to facilitate grant-free mechanisms potentially arising from URLLC and mMTC. Also, accommodating several services in terms of different subcarrier spacing with diminished guard bands in the system bandwidth becomes viable. 
The second is to lower the peak-to-average power ratio (PAPR) of baseband signals. By doing so, the input backoff (IBO) to maintain the operation in PA linear region can be decreased in order to substantially improve the PA efficiency, the battery life, and the coverage range of a UE. It also brings down the cost of hardware implementation, particularly for uplink and sidelink devices supporting very high carrier frequencies. 
A joint consideration of lowering OSBE and PAPR together is essential, since undesirable PA nonlinearity actually causes spectral regrowth that may deteriorate expected spectral containment virtue. 
Pure OFDM waveform and its orthogonal frequency division multiple access (OFDMA) usage, known to possess severe OSBE and rather high PAPR, need to be improved \cite{Lien2017_HYM,Zaidi2016_WD,Wunder2014_WD, Guan2017_WD, Zaidi2016VTC_WD}.

To deal with the two challenging yet crucial issues, auxiliary techniques such as windowing, filtering, and precoding have been proposed and studied for years \cite{You2014_WDIS_SL, Rahmatallah2013_WDIS_PAE_PAPR}, and several new waveform candidates arose since then \cite{Zhang2016_WD, Qualcomm162199_3GPPTDOC}. 
Weighted overlap-and-add (WOLA) OFDM \cite{Zayani2016_WD_WOFDM}, as a straightforward windowing approach merely for OSBE suppression, prevents steep changes between two rectangularly pulsed OFDM blocks. Universal-filtered OFDM (UF-OFDM) \cite{Schaich2014_WD_FOFDM} and filtered-OFDM (f-OFDM) \cite{Abdoli2015_WD_FOFDM} introduce the functionality of subband-wise filtering, which ideally results in extremely low OSBE but in practice causes increased PAPR with significant spectral regrowth \cite{Skyworks165035_3GPPTDOC}. The aforementioned three waveforms even give rise to inter-block interference (IBI) at the receiver, if the guard interval (GI) such as cyclic prefix (CP) and zero padding (ZP) cannot accommodate the composite delay spread of wireless channel and block extension. In contrary to the filter utilization, precoding techniques are usually helpful to PAPR reduction without imposing GI burden. One representative waveform is DFT-S-OFDM that has been adopted in 4G LTE uplink specification, in the name of single-carrier frequency division multiple access (SC-FDMA) \cite{Myung2008_TB}. Spectrally shaped SC-FDMA (SS-SC-FDMA) \cite{Benammar2013_WD_DFTSOFDM} is able to achieve a lower PAPR than that of DFT-S-OFDM at the cost of excess bandwidth and noise enhancement penalty (NEP) \cite{Kawamura2006_WD_DFTSOFDM}. The optimization of the spectrally shaping coefficients to statistically reduce the PAPR has been studied in \cite{Falconer2011_WD_DFTSOFDM, Yuen2012_WD_DFTSOFDM}. Moreover, DFT-S-OFDM based waveforms can easily yield low OSBE when few input data symbols and CP are replaced by zeros. A famous example is zero-tail (ZT) DFT-S-OFDM \cite{Berardinelli2013_WD_DFTSOFDM}, although it leads to a slightly higher PAPR than that of DFT-S-OFDM. There are also some studies to change the input zero symbols into specific coefficients for further enhancement \cite{Sahin2016_WD_DFTSOFDM, Berardinelli2016_WD_DFTSOFDM}. Generalized frequency division multiplexing (GFDM) \cite{Michailow2014_WD_GFDM}, known for its circular pulse shaping \cite{RezazadehReyhani2015_WD_GFDM}, can be viewed as a kind of precoded OFDM that flexibly constructs $K$ spectrally shaped $M$-point DFT precoders in front of OFDM modulator implemented with $KM$-point inverse fast Fourier transform (IFFT) \cite{Michailow2012_WD_GFDM}. GFDM is capable of offering lower out-of-band emission and PAPR than those of OFDM \cite{Zhang2017_WD}. Nevertheless, GFDM has difficulty in multiple access support, since its spectral shaping followed by each $M$-point DFT is basically performed on the whole $KM$ subcarriers, unless otherwise specified (e.g., raised cosine shaping with at most $2M$ coefficients and some null subcarriers are used \cite{Sharifian2016_WD_GFDM, Matthe2015_WD_GFDM}). GFDM has been proposed to be embedded in OFDM within a subband, together with incumbent 4G transmissions, to facilitate future heterogeneous access in legacy spectra\cite{Huang2016_HYM}. Apart from DFT-based precoding techniques, other precoder types found in the literature (e.g., \cite{Ma2011_WD_POFDM, Wu2016_WD_POFDM, Xu2017_WD_POFDM}) often have undesirable complexity and compatibility issues for 5G NR actualization.
Among the aforementioned waveform candidates, an optimal subband-wise waveform design to jointly reduce the OSBE and the PAPR with controllable NEP is absent.

In this paper, we propose a new circularly pulse-shaped (CPS) precoding OFDM waveform called CPS-OFDM and its optimization design to jointly reduce the OSBE and the PAPR with controllable NEP. CPS-OFDM features a generalized low-complexity DFT-based subband precoder structure, which possesses degrees of freedom (DoF) adaptively reducing the OSBE and the PAPR with controllable NEP. Similar to the implementation of GFDM \cite{Michailow2012_WD_GFDM, Chen2017_HYM}, the DoF can be treated as either a prototype vector or a characteristic matrix producing all entries of the CPS precoding matrix. To optimize the CPS precoder invariant of coming data, we consider the statistical quantities of OSBE and PAPR derived from the power spectral density (PSD) and the variance of instantaneous power (VIP), respectively. A quartic minimization problem with complex variables is accordingly formulated to find the optimal prototype vector. As the original problem is difficult to solve, we propose to convert it into a series of simple semidefinite programming (SDP) relaxation problems under the majorization-minimization (MM) algorithmic framework \cite{Sun2017_OPT_MM}. With our proof of the original objective function in matrix quadratic form being convex, the globally optimal solution of the original problem is guaranteed to be attained via MM iteration process \cite{Razaviyayn2013_OPT_MM}. In addition, we introduce a method of choosing the initial point, which must be in the feasible set of the first SDP relaxation problem. These achievements in CPS-OFDM can serve as a universal optimization framework for SS-SC-FDMA \cite{Kawamura2006_WD_DFTSOFDM, Falconer2011_WD_DFTSOFDM, Yuen2012_WD_DFTSOFDM} and GFDM \cite{Michailow2014_WD_GFDM}. It is worthy to note that lowering both OSBE and PAPR by means of such linear precoding also prevents imposing distortion on the transmitted signal, additional power consumption and real-time computation at the transmitter, and side information required by the receiver, as compared to other techniques such as clipping, cancellation carrier, constrained constellation shaping, and selective mapping enumerated in \cite{You2014_WDIS_SL,Rahmatallah2013_WDIS_PAE_PAPR}.

The remainder of this paper is organized as follows. In Section \ref{Sec:SystemMetric}, a subband-wise precoded OFDM transceiver model and its design problem are described. In Section \ref{Sec:CPSprec}, the proposed CPS-OFDM waveform is introduced. Three equivalent CPS precoder implementation methods are provided along with their complexity analysis. The closed-form expressions of PSD, VIP, and NEP are analytically derived. In Section \ref{Sec:OptSC}, the MM procedure of solving the proposed optimization problem is presented. In Section \ref{Sec:SimResult}, simulation results reveal the performance gains of applying the proposed scheme to 5G NR. Finally, Section \ref{Sec:Conclusion} gives concluding remarks and makes recommendations for future work.

\begin{figure*}[t]
\centering \centerline{
\includegraphics[width=1.0\textwidth,clip]{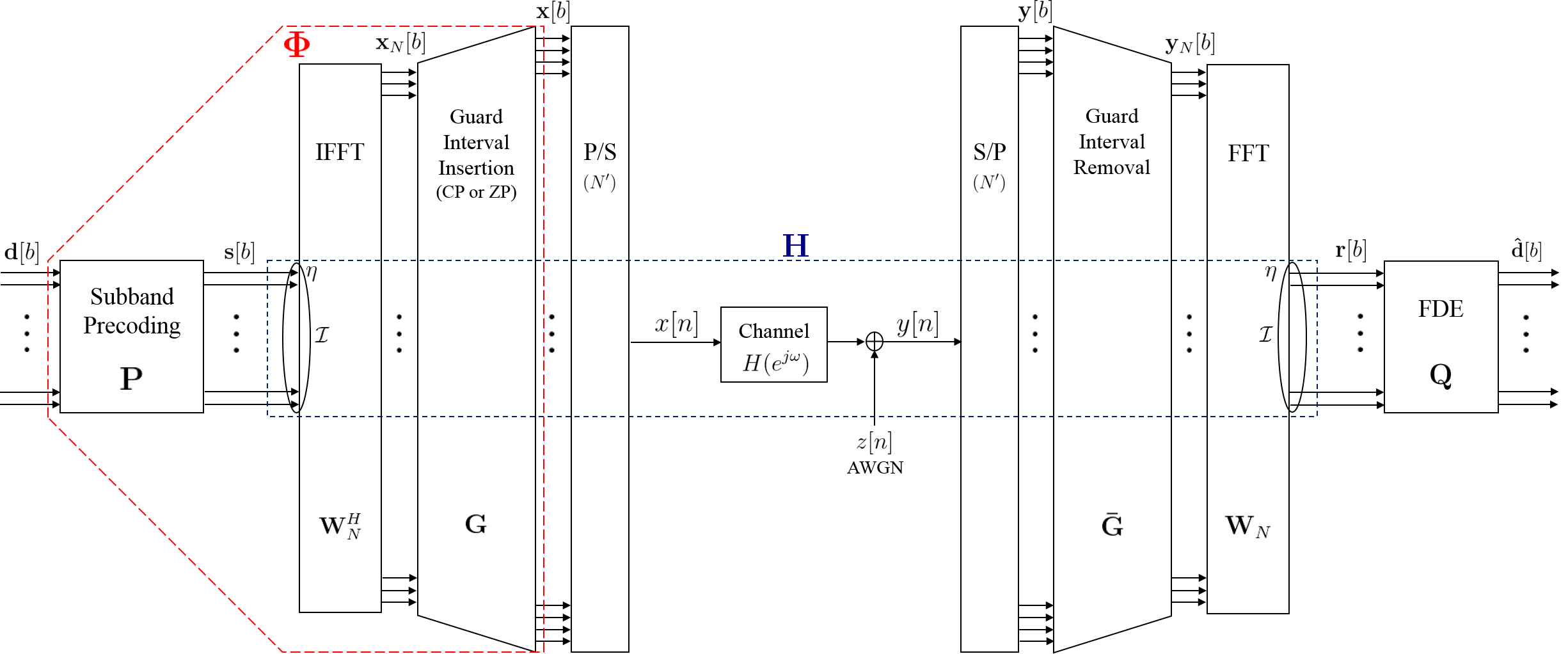}}
\caption{Subband-wise precoded-OFDM baseband transceiver model.}
\label{TxRx_PrecOFDM}
\end{figure*}

{\it Notations}: Boldfaced lower case letters such as $\mathbf{x}$ represent column vectors, boldfaced upper case letters such as $\mathbf{X}$ represent matrices, and italic letters such as $X$ represent scalars. Superscripts as in $\mathbf{X}^{T}$, $\mathbf{X}^{H}$, $\mathbf{X}^{-1}$, and $\mathbf{X}^{\circ -1}$ denote the transpose, transpose-conjugate, inverse, and Hadamard inverse operators, respectively. Calligraphic upper case letters such as $\mathcal{I}$ represent sets of discrete indices or continuous intervals. The cardinality of the discrete set $\mathcal{I}$ is described as $\left| \mathcal{I} \right|$. The submatrix of $\mathbf{X}$ formed by the column vectors with the ordered indices given in $\mathcal{I}$ is denoted by $\left[ \mathbf{X} \right] _{\mathcal{I}}$. Similarly, the subvector of $\mathbf{x}$ is denoted as $\left[ \mathbf{x} \right]_{\mathcal{I}}$. Let $\mathbf{0}_{N\times M}$, $\mathbf{1}_{M}$, $\mathbf{I}_{N}$, and $\mathrm{diag}\left( \mathbf{x} \right)$ be the $N\times M$ zero matrix, the $M\times 1$ vector of ones, the $N\times N$ identity matrix, and the diagonal matrix containing $\mathbf{x}$ on its diagonal, respectively. Functions $\mathrm{tr}\left( \mathbf{X}\right)$, $\mathrm{vec}\left( \mathbf{X}\right)$, $\mathrm{rank}\left( \mathbf{X}\right)$, and $\lambda_{\max}\left( \mathbf{X}\right)$ are the trace, the column-wise vectorization, the rank, and the largest eigenvalue of $\mathbf{X}$, respectively. Operators $\mathbb{E}\{ \cdot \}$, $\Re \{ \cdot \}$, $\left | \cdot  \right |$, $\left \| \cdot \right \|_{2}$, $\left \| \cdot \right \|_{\infty}$, $\left \langle \cdot \right \rangle _{S}$, $\circledast_{S}$, and $\otimes $ denote expectation, real part of a complex number, modulus of a complex scalar, Euclidean norm, infinity norm, modulo $S$, $S$-period circular convolution, and Kronecker product, respectively. The Kronecker delta is $\delta_{k,k'}=1$ for $k=k'$, and $\delta_{k,k'}=0$ otherwise. The expression $\mathbf{X} \succeq (\succ) \mathbf{Y}$ means that $\mathbf{X}-\mathbf{Y}$ is a positive semidefinite (definite) matrix. Throughout the paper we adopt zero-based indexing. The $N$-point normalized DFT matrix denoted by $\mathbf{W}_{N}$ is defined as that the $(k,n)$th entry of $\mathbf{W}_{N}$ is $e^{-j2\pi kn/N} / \sqrt{N}$. The $i$th entry of $\mathbf{x}$ and the $(i,j)$th entry of $\mathbf{X}$ are denoted by $[\mathbf{x}]_{i}$ and $[\mathbf{X}]_{i,j}$, respectively. Given any positive integer $N$,  $\mathcal{Z}_{N}$ stands for the set $\left\{ 0,1,\cdots,N-1 \right\}$. For any positive integers $K$ and $M$, the $M\times K$ matrix denoted by $ \mathrm{reshape}\left( \mathbf{x} , M , K \right) $ has the $(m,k)$th entry being $\left[ \mathbf{x} \right]_{kM+m}$, $\forall m\in \mathcal{Z}_{M}, k\in \mathcal{Z}_{K}$. 


\section{System Model}
\label{Sec:SystemMetric}

We consider a precoded-OFDM system equipped with $N$ OFDM subcarriers, among which each user in the system is assigned a certain contiguous subcarriers. 
The baseband uplink transceiver model of the system in the view of a single user is schematized in Fig. \ref{TxRx_PrecOFDM}.
At the transmitter, the input $S \times 1$ data vector of the $b$th block transmission $\mathbf{d}[b]$ is first precoded by an $S \times S$ precoding matrix $\mathbf{P}$ to obtain $\mathbf{s}[b]=\mathbf{P}\mathbf{d}[b]$. 
The precoder $\mathbf{P}$ can be designed to obtain desired waveform properties. 
The precoded symbols $\mathbf{s}[b]$ are then assigned to $S$ contiguous OFDM subcarriers, whose indices are given in $\mathcal{I}=\left\{ \eta, \eta+1, \cdots, \eta+S-1 \right\}$ with $\eta \geq 0$ and $\eta+S \leq N$. 
An $N$-point IFFT, characterized by $\mathbf{W}_{N}^{H}$, is used for OFDM modulation. To prevent IBI stemmed from channel delay spread, a GI is added on each time domain block signal $\mathbf{x}_{N}[b] =\left[ \mathbf{W}_{N}^{H} \right]_{\mathcal{I}}\mathbf{s}[b]$. The GI insertion can be represented by a matrix $\mathbf{G}$ chosen to be either
\begin{eqnarray}
\label{CPZP}
\mathbf{G}_{\mathrm{cp}}=\begin{bmatrix}
\mathbf{0}_{G\times (N-G)}\quad \mathbf{I}_{G} \\
\mathbf{I}_{N}
\end{bmatrix}\enskip \textrm{or} \enskip
\mathbf{G}_{\mathrm{zp}}=\begin{bmatrix}
\mathbf{I}_{N} \\
\mathbf{0}_{G\times N}
\end{bmatrix} \nonumber
\end{eqnarray}
for CP or ZP of length $G$, respectively. Thus, the transmitted signal of the $b$th block containing $N'=N+G$ samples is formulated as
\begin{eqnarray}
\label{TxSignal}
\mathbf{x}[b] = \mathbf{G} \mathbf{x}_{N}[b] =\mathbf{G} \left[ \mathbf{W}_{N}^{H} \right]_{\mathcal{I}}\mathbf{P}\mathbf{d}[b],
\end{eqnarray}
where $\mathbf{G}=\mathbf{G}_{\mathrm{cp}}$ or $\mathbf{G}=\mathbf{G}_{\mathrm{zp}}$, depending on the choice of GI. We denote $\mathbf{\Phi}=\mathbf{G} \left[ \mathbf{W}_{N}^{H} \right]_{\mathcal{I}}\mathbf{P}$ and call it a synthesis matrix. After parallel-to-serial conversion (P/S) of (\ref{TxSignal}), the digital baseband transmit signal
\begin{eqnarray}
\label{TxSeqSignal}
{x}[n]=\sum_{b=-\infty}^{\infty} \sum_{i=0}^{S-1} \left[ \mathbf{\Phi} \right]_{\left \langle n-bN' \right \rangle _{N'},i} \left[ \mathbf{d}[b] \right]_{i}
\end{eqnarray}
is sent over a frequency-selective channel. The channel can be modeled as a linear time-invariant finite impulse response filter $H \left( e^{j\omega}\right)=\sum_{l=0}^{L}h[l] e^{-j\omega l} $ with the channel order $L$. The channel impulse response vector is $\mathbf{h} = \left[ h[0] \enskip h[1] \cdots h[L] \right]^{T}$.

At the receiver, serial-to-parallel conversion (S/P) of sequentially incoming ${y}[n]= \sum_{l=0}^{L }h[l]  {x}[n-l]+{z}[n]$ is first performed, where ${z}[n]$ is a complex additive white Gaussian noise (AWGN) with variance $N_{0}$. The received signal of the $b$th block can be written as \cite{Muquet2002_WD}
\begin{eqnarray}
\label{RxSignal}
\mathbf{y}[b] = \mathbf{T}_{\mathrm{low}}\mathbf{x}[b]+ \underbrace{ \mathbf{T}_{\mathrm{up}}\mathbf{x}[b-1] }_{\textrm{IBI}}+ \mathbf{z}[b],
\end{eqnarray}
where $\mathbf{T}_{\mathrm{low}}$ is an $N' \times N' $ lower triangular Toeplitz matrix with its first column $\left[ \mathbf{h}^{T} \enskip 0 \cdots 0 \right]^{T}$, $\mathbf{T}_{\mathrm{up}}$ is an $N' \times N' $ upper triangular Toeplitz matrix with its first row  $\left[ 0 \cdots 0 \enskip h[L] \cdots h[1]  \right]$, and $\mathbf{z}[b]$ is a blocked noise vector with its covariance matrix $N_{0}\mathbf{I}_{N'}$. For CP removal or overlap-add manipulation \cite{Muquet2002_WD}, a matrix $\mathbf{\bar G}$ is chosen as
\begin{eqnarray}
\label{removeCPZP}
\mathbf{\bar G}_{\mathrm{cp}}=\begin{bmatrix}
\mathbf{0}_{N\times G}\quad \mathbf{I}_{N}
\end{bmatrix}\enskip \textrm{or} \enskip
\mathbf{\bar G}_{\mathrm{zp}}=\begin{bmatrix}
\mathbf{I}_{G} & \mathbf{0} & \mathbf{I}_{G}  \\
\mathbf{0}     & \mathbf{I}_{N-G} & \mathbf{0}
\end{bmatrix}, \nonumber
\end{eqnarray}
respectively. Under the assumption of $G\geq L$, we can extract the $N\times 1$ received vector $\mathbf{y}_{N}[b]$ from (\ref{RxSignal}) without IBI, i.e., 
\begin{eqnarray}
\label{effRxSignal}
\mathbf{y}_{N}[b] = \mathbf{\bar G} \mathbf{y}[b] =  \mathbf{\bar G}\mathbf{T}_{\mathrm{low}}\mathbf{x}[b]+  \mathbf{\bar G}\mathbf{z}[b],
\end{eqnarray}
where $\mathbf{\bar G}=\mathbf{\bar G}_{\mathrm{cp}}$ or $\mathbf{\bar G}=\mathbf{\bar G}_{\mathrm{zp}}$, depending on the type of $\mathbf{G}$ used in the transmitter. Then, an $N$-point FFT is applied to $\mathbf{y}_{N}[b]$. The received frequency-domain signal to be processed is expressed as \cite{Lin2011_TB}
\begin{eqnarray}
\label{procRxSignal}
\mathbf{r}[b] =  \mathbf{H} \mathbf{P}\mathbf{d}[b]+\mathbf{v}[b],
\end{eqnarray}
where $\mathbf{H}= \mathrm{diag}\left( \left[ \mathbf{W}_{N} \left[ \mathbf{h}^{T} \enskip \mathbf{0}_{(N-L-1)\times 1}^{T} \right]^{T} \right]_{\mathcal{I}} \right)$ has diagonal elements corresponding to channel frequency response on the occupied $S$ subcarriers and $\mathbf{v}[b]=\left[ \mathbf{W}_{N}^{T} \right]_{\mathcal{I}}^{T}\mathbf{\bar G}\mathbf{z}[b]$ is the noise vector. Finally, the received data vector can be linearly obtained by
\begin{eqnarray}
\label{RxdataVec}
\mathbf{\hat d}[b] = \mathbf{Q}\mathbf{r}[b],
\end{eqnarray}
where $\mathbf{Q}$ is a frequency domain equalization (FDE) matrix in zero forcing (ZF) or minimum mean square error (MMSE) sense.

In some cases, the data vector $\mathbf{d}[b]$ is composed of $D$ data symbols drawn from a quadrature amplitude modulation (QAM) constellation and $Z$ zeros, $S=D+Z$. Let $\mathcal{D}\subseteq \mathcal{Z}_{S}$ be the set of indices indicating the locations of $D$ data symbols in $\mathbf{d}[b]$, $\forall b$. All data symbols are assumed to be zero-mean, independent and identically distributed (i.i.d.) with symbol power $E_{\mathrm{s}}$, i.e.,
\begin{eqnarray}
\label{zeromeanData}
\mathbb{E}\left\{ \left[ \mathbf{d}[b] \right]_{\mathcal{D}} \right\}=\mathbf{0}_{D\times 1},\enskip \forall b,
\end{eqnarray}
\begin{eqnarray}
\label{iidData}
\mathbb{E}\left\{ \left[ \mathbf{d}[b] \right]_{\mathcal{D}} \left[ \mathbf{d}[b'] \right]_{\mathcal{D}}^{H}  \right\} = E_{\mathrm{s}}\mathbf{I}_{D} \delta_{b,b'}, \enskip \forall b,b' ,
\end{eqnarray}
and jointly wide sense stationary and uncorrelated with noise. Let $ \mathbf{\bar P} = \left[ \mathbf{P} \right]_{\mathcal{D}}$. The matrix $\mathbf{Q}$ in (\ref{RxdataVec}) can be chosen as \cite{Lin2011_TB}
\begin{eqnarray}
\label{ZFFDEMx}
\mathbf{Q}_{\mathrm{ZF}} = \left[ \left( \mathbf{H}  \mathbf{\bar P} \right)^{H}\left( \mathbf{H}  \mathbf{\bar P} \right) \right]^{-1}\left( \mathbf{H}  \mathbf{\bar P} \right)^{H}
\end{eqnarray}
or
\begin{eqnarray}
\label{MMSEFDEMx}
\mathbf{Q}_{\mathrm{MMSE}} = \left[ \left( \mathbf{H}  \mathbf{\bar P} \right)^{H}\left( \mathbf{H}  \mathbf{\bar P} \right)+\frac{N_{0}}{E_{\mathrm{s}}} \mathbf{I}_{D}\right]^{-1}\left( \mathbf{H}  \mathbf{\bar P} \right)^{H}
\end{eqnarray}
in the sense of ZF-FDE or MMSE-FDE, respectively.


\subsection{Quantifying Spectral Sidelobe Leakage Using OSBEP}
\label{subSec:OSBE}
Spectral sidelobe leakage of a user, referred to as OSBE, is evaluated by calculating the PSD. Under the assumptions of (\ref{zeromeanData}) and (\ref{iidData}), the PSD of (\ref{TxSeqSignal}) is given by \cite{Lin2011_TB}
\begin{eqnarray}
\label{PSD}
{S}_{ x}\left( e^{j\omega} \right)=\frac{E_{\mathrm{s}}}{N'} \sum_{i\in \mathcal{D}} \left| \Phi_{i}\left( e^{j\omega } \right)  \right|^{2},\enskip \omega \in [ -\pi,\pi ), 
\end{eqnarray}
where $\Phi_{i}\left( e^{j\omega } \right) = \sum_{n'=0}^{N'-1} \left[ \mathbf{\Phi} \right]_{n',i} e^{-j\omega n}$ represents the synthesis filter used for the $i$th data stream (i.e., $\left[ \mathbf{d}[b] \right]_{i}$, $\forall b$). Denote $\mathcal{F}_{\mathrm{OSB}}\subset [ -\pi,\pi )$ as an OSB region to be concerned. To quantify the OSBE, a commonly used approach is to compute its total power (OSBEP)\footnote{It is worthy to note that an actual baseband PSD expression involves an interpolation filter used in a digital-to-analog converter (DAC). Specifically, the analog transmitted signal ${x}(t)$ is obtained by passing ${x}[n]$ through a DAC with a sampling period $T_{\mathrm{s}}$ and an interpolation filter $G(f)$. The PSD of ${x}(t)$ is ${\tilde S}_{x}\left( f \right) =\frac{1}{T_{\mathrm{s}}} {S}_{x}\left( e^{j f T_{\mathrm{s}} } \right) \left| G(f) \right|^{2} $ \cite{Lin2011_TB}. Since $G(f)$ mainly affects the spectral attenuation outside the sampling bandwidth $[ -\pi /T_{\mathrm{s}},\pi /T_{\mathrm{s}} )$, for simplicity we assume $\left| G(f) \right|=1$ for $|f|<\pi/T_{\mathrm{s}}$ and $\left| G(f) \right| =0$ otherwise, and address the amount of OSBE by (\ref{OSBEP}) in this study.} \cite{Ma2011_WD_POFDM, Wu2016_WD_POFDM, Xu2017_WD_POFDM}
\begin{eqnarray}
\label{OSBEP}
\gamma_{x} = \int_{\omega\in \mathcal{F}_{\mathrm{OSB}}}  S_{ x}\left( e^{j\omega } \right)  \frac{d \omega}{2\pi}.
\end{eqnarray}


\subsection{VIP to Measure PA Efficiency}
A measure closely related to the nonlinear distortion caused by PA, namely, VIP, is known as a more practical metric than PAPR \cite{Behravan2006_WDIS_PAE}. The main reason is that keeping power efficiency sufficiently high is usually much more important to UEs than taking large IBO \cite{Behravan2009_WDIS_PAE_VIP}. The VIP averaged over the $b$th block is defined as \cite{Falconer2011_WD_DFTSOFDM, Yuen2012_WD_DFTSOFDM, Sharifian2016_WD_GFDM, Wu2016_WD_POFDM}
\begin{eqnarray}
\label{DefVIP}
{\bar \sigma}_{x}^{2}   &=& \frac{1}{N}  \sum_{n=0}^{N-1} \mathbb{E} \left\{ \left[ \left| x_{n}[b] \right|^{2} - {\bar \mu}_{x}  \right]^{2} \right\} \nonumber \\
&=&
\frac{1}{N}  \sum_{n=0}^{N-1} \mathbb{E} \left\{ \left| x_{n}[b] \right|^{4} \right\} - {\bar \mu}_{x}^{2},
\end{eqnarray} 
where $x_{n}[b]=\left[ \mathbf{x}_{N}[b] \right]_{n}$ for $n \in \mathcal{Z}_{N}$ and ${\bar \mu}_{x}$ is the mean of instantaneous power averaged over the $b$th block given by 
\begin{eqnarray}
\label{DefMIP}
{\bar \mu}_{x} =\frac{1}{N} \sum_{n=0}^{N-1} \mathbb{E} \left\{ \left| x_{n}[b] \right|^{2} \right\}.
\end{eqnarray}
With (\ref{zeromeanData}) and (\ref{iidData}), we further define $\sigma_{\mathrm{d}}^{4}=\mathbb{E}\left\{ \left| \left[ \mathbf{d}[b] \right]_{i} \right|^{4}  \right\}$, $\forall i \in \mathcal{D}$, $\forall b$. Then, it can be easily found that (\ref{DefVIP}) and (\ref{DefMIP}) are independent of the block index $b$.

In subsequent contents, the block index ``$[b]$'' is omitted for notational brevity, since the design of $\mathbf{P}$ taking (\ref{PSD})-(\ref{DefMIP}) into account is invariant of coming data and there is no IBI in (\ref{effRxSignal}).



\subsection{Subband Precoder Design Problem}
\label{subSec:PrecoderProblem}
This paper studies the subband precoder design problem for the precoded-OFDM system described in Section \ref{Sec:SystemMetric}.
Specifically, we intend to design the $S\times S$ precoding matrix $\mathbf{P}$ such that the OSBEP (\ref{OSBEP}) and the VIP (\ref{DefVIP}) can be simultaneously reduced as compared to OFDMA. The detailed problem statement will be formulated in the Sections \ref{subSec:CPSprec_PSDOSBEP}, \ref{subSec:CPSprec_VIP}, and \ref{Sec:OptSC}.

In addition, it is also desirable for the precoding matrix ${\mathbf P}$ to possess a low-complexity implementation and at the same time not to cause significant receiver performance degradation.
For the complexity concern, we notice that in the most general form, there are $S^{2}$ complex-valued coefficients in ${\bf P}$ to be specified, resulting in an undesired quadratic order complexity (i.e., ${\cal O}(S^2)$). 
We thus seek some constraints on the precoder structure to make the implementation efficient in linearithmic order. 
On the other hand, for the concern of performance degradation, it will be helpful when the precoder $\mathbf{P}$ is chosen to be unitary (i.e., $\mathbf{P}^{H}\mathbf{P}=\mathbf{I}_{S}$), so that the NEP at the receiver will be avoided. 
The above two issues will be addressed in more details in Sections \ref{subSec:CPSprec_ImplementComplexity} and \ref{subSec:CPSprec_NEP}, respectively.

\section{Proposed Circularly Pulse-Shaped (CPS) Precoding Method for OFDM}
\label{Sec:CPSprec}
To tackle the problem stated in Section \ref{subSec:PrecoderProblem}, we propose a circularly pulse-shaped (CPS) precoding method for OFDM in this section. 
Specifically, the CPS precoding matrix is designed as
\begin{eqnarray}
\label{CPSPrec_direct}
\mathbf{P}=\mathbf{W}_{S}\mathbf{A},
\end{eqnarray}
where $\mathbf{W}_{S}$ is an $S$-point normalized DFT matrix and $\mathbf{A}$ is an $S\times S$ GFDM matrix with the $(kM+m)$th column vector $\mathbf{a}_{k,m}$ derived from an $S\times 1$ prototype pulse vector $\mathbf{a}_{0,0}$, i.e., 
\begin{eqnarray}
\label{a_vec}
[\mathbf{a}_{k,m}]_{s}=[\mathbf{a}_{0,0}]_{ {\left \langle s-mK \right \rangle _{S}} } e^{j2\pi \frac{k}{K}s}, 
\end{eqnarray}
$\forall s \in \mathcal{Z}_{S}$, $k \in \mathcal{Z}_{K}$, $m \in \mathcal{Z}_{M}$, $S=KM$ \cite{Michailow2014_WD_GFDM}. 
The subband-wise transmitted signal (\ref{TxSignal}) is then reformulated as 
\begin{eqnarray}
\label{TxSignal_CPSOFDM}
\mathbf{x} = \mathbf{G} \mathbf{x}_{N} = \mathbf{G} \left[ \mathbf{W}_{N}^{H} \right]_{\mathcal{I}} \mathbf{W}_{S}\mathbf{A} \mathbf{d},
\end{eqnarray}
which can inherit the desired properties of spectral containment and reduced PAPR from the GFDM. This proposed waveform, called CPS-OFDM, endows every user with flexibility to determine its own circular pulse shaping in terms of $K$, $M$, and $\mathbf{a}_{0,0}$, while preserving frequency domain orthogonality with other users in the system.

In the following subsections, we will elaborate on the precoder implementation, the OSBEP, the VIP, and the NEP of CPS-OFDM. Based on these analytic derivations, we are able to formulate an optimization problem in the next section.


\subsection{CPS Precoder Implementation and Complexity Analysis}
\label{subSec:CPSprec_ImplementComplexity}
Three equivalent CPS precoder implementation methods and their complexity are studied in this subsection. The first is a direct implementation done by the matrix multiplication in (\ref{CPSPrec_direct}). The second is referred to as a frequency domain implementation with the prototype shaping vector $\mathbf{p}=\mathbf{W}_{S}\mathbf{a}_{0,0}$. The third is called a characteristic matrix domain implementation with the characteristic matrix $\mathbf{\Gamma}=\sqrt{S/ \rho} \enskip \mathrm{reshape}\left( \mathbf{p} , M,K \right) \mathbf{W}_{K}^{H}$, where $\rho =  \left\| \mathbf{p} \right\|_{2}^{2} $ denotes the energy of $\mathbf{p}$. Note that $\mathbf{a}_{0,0}$, $\mathbf{p}$, and $\mathbf{\Gamma}$ are invertible linear transformations of one another.

In the frequency domain implementation, we consider $\mathbf{P}$ composed of $K$ submatrices, i.e., $\mathbf{P} = \left[ \mathbf{P}_{0} | \mathbf{P}_{1} | \cdots | \mathbf{P}_{K-1}  \right]$, where $\mathbf{P}_{k} = \left[ \mathbf{P} \right]_{\mathcal{I}_{k}}$, $\mathcal{I}_{k}= \left\{ kM, kM+1, \cdots , kM+M-1 \right\}$. According to (\ref{a_vec}), the $k$th submatrix of $\mathbf{A}$, defined as $\mathbf{A}_{k} \triangleq \left[ \mathbf{A} \right]_{\mathcal{I}_{k}}$, can be expressed as \cite{Michailow2012_WD_GFDM}
\begin{eqnarray}
\label{subAmatrix}
\mathbf{A}_{k} \triangleq \left[ \mathbf{A} \right]_{\mathcal{I}_{k}} =\mathbf{W}_{S}^{H} \mathbf{C}_{kM} \mathrm{diag}\left( \mathbf{p} \right) \mathbf{R} \mathbf{W}_{M},
\end{eqnarray}
where $\mathbf{R}= \mathbf{1}_{K} \otimes \mathbf{I}_{M} $ is a repetition matrix and $\mathbf{C}_{kM}$ is a downshift permutation matrix defined as
\begin{eqnarray}
\label{Cmatrix}
\mathbf{C}_{0}=\mathbf{I}_{S}, \enskip
\mathbf{C}_{kM}=\begin{bmatrix}
\mathbf{0} & \mathbf{I}_{kM} \\
\mathbf{I}_{S-kM} & \mathbf{0}
\end{bmatrix}. \nonumber
\end{eqnarray}
Hence, the $k$th submatrix of $\mathbf{P}$ is obtained by
\begin{eqnarray}
\label{subPmatrix}
\mathbf{P}_{k}=\mathbf{W}_{S}\mathbf{A}_{k}=\mathbf{C}_{kM} \mathrm{diag}\left( \mathbf{p} \right)  \mathbf{R} \mathbf{W}_{M}^{},
\end{eqnarray}
which can be interpreted as the spectrally shaped $M$-point DFT precoder applied to $\mathbf{d}_{k}= \left[ \mathbf{d} \right]_{\mathcal{I}_{k}}$. 

In the characteristic matrix domain implementation, we refer to the results in \cite[Lemma 1(c)]{Chen2017_HYM} for $\mathbf{A}$, and by analogy derive
\begin{eqnarray}
\mathbf{P} = \left( \mathbf{W}_{K} \otimes \mathbf{I}_{M} \right) \mathrm{diag} \left( \mathrm{vec}\left( \mathbf{\Gamma}\right) \right) \left( \mathbf{W}_{K}^{H} \otimes \mathbf{W}_{M} \right).
\end{eqnarray}
With this form, the unitarity of the CPS precoding matrix can be easily identified, i.e., $\mathbf{P}$ is unitary if and only if $ | \left[ \mathbf{\Gamma} \right]_{m,k} |=1$, $\forall k\in \mathcal{Z}_{K},m\in \mathcal{Z}_{M}$. Interested readers may refer to \cite[Theorem 1]{Chen2017_HYM} for more details about unitary GFDM matrices.

\begin{figure}[t]
\centering
   \subfigure[]{\includegraphics[width=0.3\textwidth,clip]{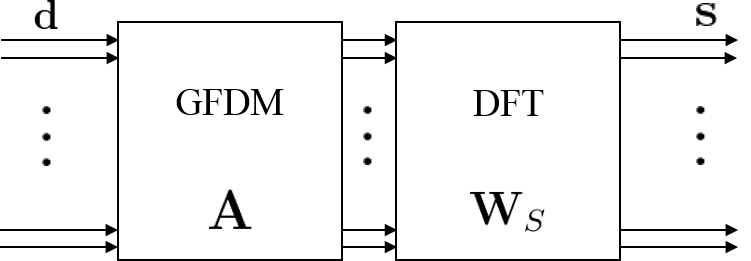}} \\
   \subfigure[]{\includegraphics[width=0.5\textwidth,clip]{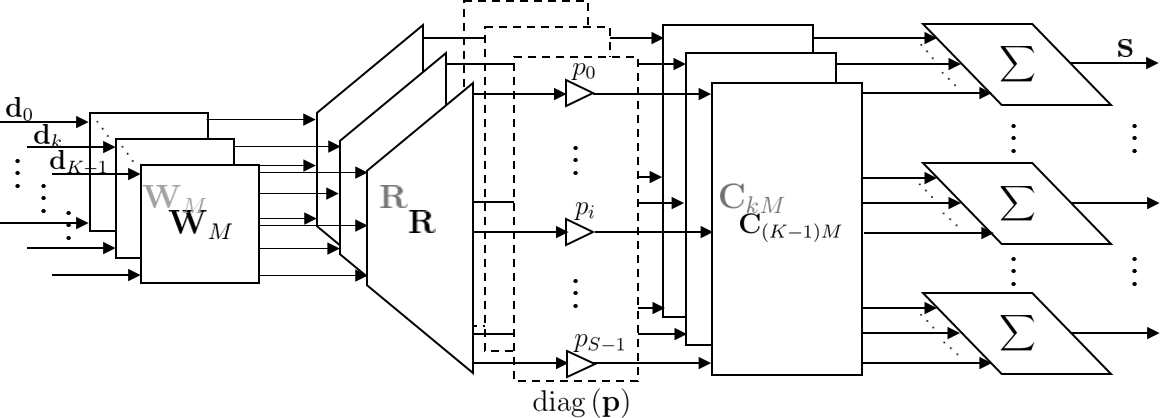}} \\
   \subfigure[]{\includegraphics[width=0.5\textwidth,clip]{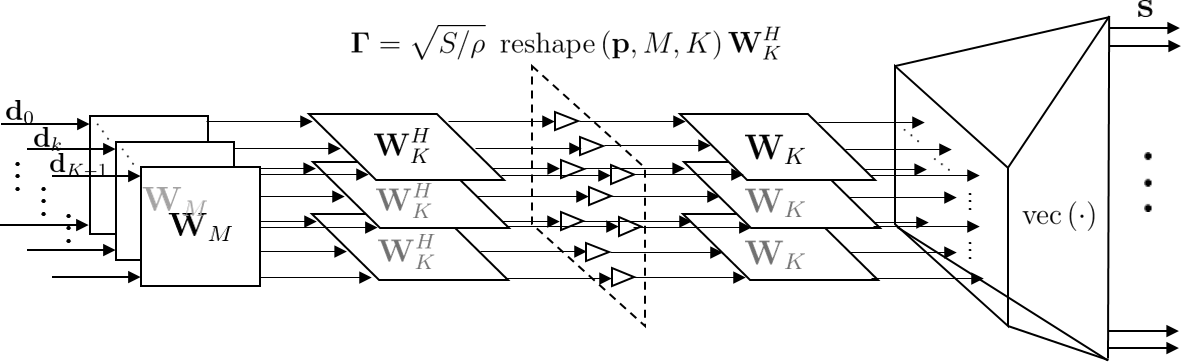}}
   \caption{Propsoed CPS precoder with (a) direct implementation (\ref{PrecSym1}), (b) frequency domain implementation (\ref{PrecSym2}), and (c) characteristic matrix domain implementation (\ref{PrecSym3}).}
   \label{CPSprecImplement}
\end{figure}

In summary, the precoded symbols on the OFDM subcarriers indexed by $\mathcal{I}$ can be produced with
\begin{subequations}
\begin{align}
\mathbf{s}
&\ \label{PrecSym1}
= \mathbf{W}_{S} \mathbf{A} \mathbf{d} \\
&\ \label{PrecSym2}
=  \sum_{k=0}^{K-1} \mathbf{C}_{kM} \mathrm{diag}\left( \mathbf{p} \right)  \mathbf{R} \mathbf{W}_{M} \mathbf{d}_{k} \\
&\ \label{PrecSym3}
= \left( \mathbf{W}_{K} \otimes \mathbf{I}_{M} \right) \mathrm{diag} \left( \mathrm{vec}\left( \mathbf{\Gamma}\right) \right) \left( \mathbf{W}_{K}^{H} \otimes \mathbf{W}_{M} \right) \mathbf{d},
\end{align}
\end{subequations}
whose building blocks are displayed in Fig. \ref{CPSprecImplement}(a), Fig. \ref{CPSprecImplement}(b), and Fig. \ref{CPSprecImplement}(c), respectively.

The precoder complexity with respect to different implementation methods are evaluated by computing the number of complex multiplications (CMs) required to send $KM$ data symbols. The direct implementation (\ref{PrecSym1}) inefficiently takes $KM \left( KM+ \log_{2}KM \right)$ CMs. The frequency domain implementation (\ref{PrecSym2}) needs $KM \left( \log_{2}M + K\right)$ CMs, of which an $M$-point DFT and an element-wise multiplication of $\mathbf{p}$ contain $M\log_{2}M$ and $KM$ CMs, respectively. Note that the repetition and the permutation operations are without multiplier units. The characteristic matrix domain implementation (\ref{PrecSym3}) uses $KM\log_{2}M+MK\log_{2}K+KM+MK\log_{2}K=KM\left( \log_{2}K^{2}M+1\right)$ CMs, which are more than the required number of CMs of (\ref{PrecSym2}) only when $\frac{K-1}{2} < \log_{2}K$. Compared to the $S$-point DFT precoder used for DFT-S-OFDM, the CPS precoder realized in (\ref{PrecSym2}) and (\ref{PrecSym3}) increase the complexity by $KM(K-\log_{2}K)$ and $KM(1+\log_{2}K)$ CMs, respectively, because of its additional $S$ DoF and scalability in $K$. Benefited from the characteristic matrix domain implementation method, the proposed CPS precoder (\ref{CPSPrec_direct}) can be efficiently realized with the complexity in the linearithmic order ${\cal O}(KM \log_{2}KM)$.

\subsection{PSD and OSBEP of CPS-OFDM Transmission}
\label{subSec:CPSprec_PSDOSBEP}
The closed-form expressions of the PSD and the OSBEP of CPS-OFDM are analytically derived in terms of the prototype shaping vector $\mathbf{p}$ in this subsection. Using (\ref{subPmatrix}) and assuming $\mathbf{G}=\mathbf{G}_{\mathrm{cp}}$ whose CP length is now denoted by $G'$, every entry of the synthesis matrix $\mathbf{\Phi}$ is obtained by
\begin{eqnarray}
\label{PhiMx_entry}
\left[ \mathbf{\Phi} \right]_{n',s} = \frac{1}{\sqrt{NM}} \sum_{i=0}^{S-1} \left[ \mathbf{C}_{kM} \mathbf{p}  \right]_{i} e^{-j \frac{2\pi}{M}im} e^{j \frac{2\pi}{N}(\eta+i)(n'-G')} 
\end{eqnarray}
$\forall n' \in \mathcal{Z}_{N'}$, $s=kM+m  \in \mathcal{Z}_{S}$, $k \in \mathcal{Z}_{K}$, $m \in \mathcal{Z}_{M}$. The square-magnitude response of the $(kM+m)$th synthesis filter can then be described as
\begin{eqnarray}
\label{sqrmagSynthesis}
\left| {\Phi}_{k,m} \left( e^{j\omega }\right) \right|^{2} = \left| \mathbf{w}_{m}^{H}\left( e^{j\omega } \right)  \mathbf{C}_{kM} \mathbf{p} \right|^{2} ,
\end{eqnarray}
where
\begin{eqnarray} 
\left[ \mathbf{w}_{m}\left( e^{j\omega } \right) \right]_{i} =\frac{1}{\sqrt{NM}}e^{j 2\pi i \left( \frac{m}{M}+\frac{G'}{N} \right) }W^{*}\left( e^{j \left[  \omega-\frac{2\pi}{N}(\eta+i) \right]} \right),  \nonumber \\
W\left( e^{j \left[  \omega-\frac{2\pi}{N}(\eta+i) \right]} \right)=\sum_{n'=0}^{N+G'-1} e^{-j \left[ \omega - \frac{2\pi}{N} (\eta +i)\right]n'} ,\enskip \forall i \in \mathcal{Z}_{S}. \nonumber 
\end{eqnarray}
Let $\mathcal{K} \subseteq \mathcal{Z}_{K}$ and $\mathcal{M} \subseteq \mathcal{Z}_{M}$ be two sets of indices corresponding to the positions of $D$ data symbols in $\mathbf{d}$, $D=\left| \mathcal{K} \right| \left| \mathcal{M} \right| \leq S$.
Substituting (\ref{sqrmagSynthesis}) into (\ref{PSD}), the PSD of CPS-OFDM transmission is 
\begin{eqnarray}
\label{PSD_CPSOFDM}
{S}_{ x}\left( e^{j\omega} \right) = \frac{E_{\mathrm{s}}}{N'} \sum_{\scriptstyle k\in \mathcal{K} \atop \scriptstyle m\in \mathcal{M}}  \left| \Phi_{k,m}\left( e^{j\omega } \right)  \right|^{2} = \mathbf{p}^{H} \mathbf{\Psi} \left( e^{j\omega } \right) \mathbf{p},
\end{eqnarray}
where
\begin{eqnarray}\mathbf{\Psi} \left( e^{j\omega } \right) = \frac{E_{\mathrm{s}}}{N'}  \sum\limits_{\scriptstyle k\in \mathcal{K} \atop \scriptstyle m\in \mathcal{M}}  \mathbf{C}^{T}_{kM} \mathbf{w}^{}_{m}\left( e^{j\omega } \right) \mathbf{w}_{m}^{H}\left( e^{j\omega } \right) \mathbf{C}^{}_{kM}. \nonumber
\end{eqnarray}
Thus, the OSBEP is given by 
\begin{eqnarray}
\label{OSBEP_CPSOFDM}
\gamma_{ x} \left( \mathbf{p} \right)=  \mathbf{p}^{H} \mathbf{\Omega} \mathbf{p}, 
\end{eqnarray}
where $ \mathbf{\Omega} = \int_{\omega\in \mathcal{F}_{\mathrm{OSB}}}  \mathbf{\Psi} \left( e^{j\omega } \right)  \frac{d \omega}{2\pi} \succ \mathbf{0}_{S\times S}$. For $\mathbf{G}=\mathbf{G}_{\mathrm{zp}}$ and $\mathbf{G}=\mathbf{I}_{N}$, setting $G'=0$ in (\ref{PhiMx_entry}) will result in the corresponding alternative expressions of (\ref{sqrmagSynthesis})-(\ref{OSBEP_CPSOFDM}).


\subsection{VIP of CPS-OFDM Signals}
\label{subSec:CPSprec_VIP}
\label{subSec:VIPofCPSOFDM}
The VIP function (\ref{DefVIP}) to quantify the envelope fluctuations of CPS-OFDM signals is given in this subsection. According to (\ref{PhiMx_entry}) with $G'=0$, the $n$th time domain sample of the block signal $\mathbf{x}_{N}$ in (\ref{TxSignal_CPSOFDM}) can be expressed as
\begin{eqnarray}
\label{xnsample}
x_{n}  = e^{j\frac{2\pi}{N}\eta n} \sum_{\scriptstyle k\in \mathcal{K} \atop \scriptstyle m\in \mathcal{M}} d^{}_{k,m} \mathbf{e}_{m,n}^{H} \mathbf{C}^{}_{kM} \mathbf{p},
\end{eqnarray}
where $d_{k,m}= \left[ \mathbf{d}_{k} \right]_{m}$, $\left[ \mathbf{e}_{m,n} \right]_{i} = \frac{1}{\sqrt{NM}} e^{j2\pi i \left( \frac{m}{M} - \frac{n}{N} \right)}$, $\forall i \in \mathcal{Z}_{S}$. Based on (\ref{iidData}), (\ref{DefMIP}), and (\ref{xnsample}), the mean of instantaneous power of CPS-OFDM signals
\begin{eqnarray}
\label{MIP_CPSOFDM}
{\bar \mu}_{x} =\frac{1}{N} \sum_{n=0}^{N-1} \mathbb{E} \left\{ \left| x_{n} \right|^{2} \right\} =  \frac{ \left|  \mathcal{K} \right| \left|  \mathcal{M} \right|  E_{\mathrm{s}}}{NM}  \left\| \mathbf{p} \right\|_{2}^{2} = \frac{DE_{\mathrm{s}}}{NM} \rho
\end{eqnarray}
is a constant due to $\sum_{n=0}^{N-1} \mathbf{e}^{}_{m,n}  \mathbf{e}_{m,n}^{H} = \frac{1}{M}\mathbf{I}_{S} $, $\mathbf{C}^{T}_{kM}\mathbf{C}^{}_{kM} = \mathbf{I}_{S}  $, and $\mathbf{p}$ being fixed energy $\rho$. Hence, the VIP of CPS-OFDM signals ${\bar \sigma}_{x}^{2} ( \mathbf{p} ) =  f(\mathbf{p}) - {\bar \mu}_{x}^{2} $ can be sufficiently described by the quartic function $ f(\mathbf{p}) = \frac{1}{N}  \sum_{n=0}^{N-1} \mathbb{E} \left\{ \left| x_{n} \right|^{4} \right\} $, of which
\begin{align}
\label{quartic_xn}
& \mathbb{E}  \left\{ \left| x_{n} \right|^{4} \right\}  = \sigma_{\mathrm{d}}^{4} \sum_{\scriptstyle k\in \mathcal{K} \atop \scriptstyle m\in \mathcal{M}} \left| \mathbf{e}_{m,n}^{H} \mathbf{C}^{}_{kM}  \mathbf{p} \right|^4 +  \nonumber \\
& 2 E_{\mathrm{s}}^2 \sum_{ \scriptstyle k\in \mathcal{K} \atop \scriptstyle  m\in \mathcal{M}} \left| \mathbf{e}_{m,n}^{H} \mathbf{C}_{kM}^{}  \mathbf{p} \right|^2 \sum_{ \substack{ k'\in \mathcal{K} \\  m'\in \mathcal{M} \\ (k',m') \neq (k,m) }} \left| \mathbf{e}_{m',n}^{H} \mathbf{C}_{k'M}^{}  \mathbf{p} \right|^2
\end{align}
is derived with (\ref{zeromeanData}), (\ref{iidData}), and $\sigma_{\mathrm{d}}^{4}=\mathbb{E}\left\{ \left| d_{k,m} \right|^{4}  \right\}$.

\subsection{NEP for CPS-OFDM Signal Reception}
\label{subSec:CPSprec_NEP}
The NEP resulted from the proposed CPS precoder (\ref{CPSPrec_direct}) is formulated in this subsection. From the aspect of the characteristic matrix $\mathbf{\Gamma}=\sqrt{S/ \rho} \enskip \mathrm{reshape}\left( \mathbf{p} , M,K \right) \mathbf{W}_{K}^{H}$, it is straightforward to write the NEP in ZF sense as \cite{Chen2017_HYM}, \cite{Chen2017_WD_GFDM}
\begin{eqnarray}
\label{DefNEP}
\frac{1}{S} \sum_{k=0}^{K-1}\sum_{m=0}^{M-1}  \frac{1}{ | \left[ \mathbf{\Gamma} \right]_{m,k} |^{2}}  \leq 1+\epsilon,
\end{eqnarray}
where $\epsilon \geq 0$ is a noise tolerance factor. When $\epsilon = 0$, the CPS precoding matrix $\mathbf{P}$ is unitary. Note that $\mathbf{P}$ is invertible if and only if $\mathbf{\Gamma}$ has no zero entries \cite[Theorem 2]{Chen2017_HYM}. By the following expansion $\mathrm{vec}\left( \mathbf{\Gamma}\right) = \sqrt{S/ \rho} \enskip \mathrm{vec}\left( \left[ \mathbf{p}_{0},\mathbf{p}_{1} \cdots \mathbf{p}_{K-1} \right] \mathbf{W}_{K}^{H}\right) = \sqrt{S/ \rho}  \left(  \mathbf{W}_{K}^{H} \otimes \mathbf{I}_{M} \right)  \mathbf{p}$, we can rewrite (\ref{DefNEP}) as 
\begin{align}
\label{NEP_CPSOFDM}
\zeta(\mathbf{p}) &=  \mathrm{tr}\left(   \left[  \left(  \mathbf{W}_{K}^{H} \otimes \mathbf{I}_{M} \right)  \mathbf{p}\mathbf{p}^{H}  \left(  \mathbf{W}_{K}^{H} \otimes \mathbf{I}_{M} \right)^{H}  \right]^{\circ -1 }   \right) \nonumber \\
&\leq   (1+\epsilon) \frac{S^2}{\rho}.
\end{align}

\subsection{CPS-OFDM as A Generalized DFT-S-OFDM}
\label{subSec:CPSprec_Generalized}
CPS-OFDM is featured by its DFT-based subband precoder flexibility as shown in Fig. \ref{CPSprecImplement}(b). By properly choosing the parameters $K$, $M$, $\mathcal{K}$, $\mathcal{M}$, $\mathbf{p}$, and $\mathbf{G}$, the derivations in this section are applicable to OFDMA ($M=1$, $\mathcal{K}=\mathcal{Z}_{K}$, $\mathbf{p}= [1\enskip \mathbf{0}_{1\times (K-1)} ]^{T}$, $\mathbf{G}=\mathbf{G}_{\mathrm{cp}}$), SC-FDMA ($K=1$, $\mathcal{M}=\mathcal{Z}_{M}$, $\mathbf{p}=\mathbf{1}_{M}$, $\mathbf{G}=\mathbf{G}_{\mathrm{cp}}$), SS-SC-FDMA ($K=2$, $|\mathcal{K}|=1$, $\mathcal{M}=\mathcal{Z}_{M}$, $\mathbf{p}$ is a $2M$ root raise cosine vector, $\mathbf{G}=\mathbf{G}_{\mathrm{cp}}$) \cite{Benammar2013_WD_DFTSOFDM}, ZT DFT-S-OFDM ($K=1$, $|\mathcal{M}| < M$, $\mathbf{p}=\mathbf{1}_{M}$, $\mathbf{G}=\mathbf{I}_{N}$) \cite{Berardinelli2013_WD_DFTSOFDM}, and GFDM ($N=S=KM$, $\mathbf{G}=\mathbf{G}_{\mathrm{cp}}$) \cite{Michailow2012_WD_GFDM} as well. Different from GFDM, the proposed CPS-OFDM enables subband-wise user-specific circular pulse shaping with orthogonal multiple access support.


\section{Optimization of Prototype Shaping Vector\\ via Majorization-Minimization}
\label{Sec:OptSC}
In this section, we propose a quartic minimization problem with complex variables to find the optimal prototype shaping vector of the CPS precoder that results in low OSBE and low PAPR with controllable NEP. We also propose to solve the problem under the majorization-minimization (MM) algorithmic framework \cite{Sun2017_OPT_MM}.

The proposed optimal prototype shaping vector design problem is formulated as
\begin{subequations}
\label{OptShapingVecProblem}
\begin{align}
\underset{\mathbf{p}}{\text{minimize}} \enskip 
&\ \label{ObjFun_VIP_pvec}
{\bar \sigma}_{x}^{2} ( \mathbf{p} ) \enskip  \ldots (\textrm{VIP}) \\
\text{subject to} \enskip 
&\  \label{Const_OSBEP_pvec}
\gamma_{ x} \left( \mathbf{p} \right)  \leq U \enskip  \ldots (\textrm{OSBEP})  \\
&\  \label{Const_NEP_pvec}
\zeta \left( \mathbf{p} \right) \leq  (1+\epsilon) \frac{S^2}{\rho} \enskip  \ldots (\textrm{NEP})  \\
&\  \label{Const_FixEnergy_pvec}
 \left\| \mathbf{p} \right\|_{2}^{2} = \rho \enskip  \ldots (\textrm{Fixed energy}).
\end{align}
\end{subequations}
The objective (\ref{ObjFun_VIP_pvec}) is to minimize the VIP of baseband CPS-OFDM signals so as to reduce the impact from PA nonlinearity. However, the VIP function ${\bar \sigma}_{x}^{2} ( \mathbf{p} ) $ is a fourth-order polynomial of $\mathbf{p}$ as given by (\ref{quartic_xn}), which generally makes the problem NP-hard and difficult to be analyzed \cite{Jiang_OPT}. We will overcome this difficulty later in this section. 
The constraint (\ref{Const_OSBEP_pvec}) represents that the OSBEP (\ref{OSBEP_CPSOFDM}) of baseband CPS-OFDM transmission must be less than an upper bound $U$.
The choice of $U$ has to be a positive number no less than the minimum value $U_{\min}$ that guarantees the existence of $\mathbf{p}$ in (\ref{Const_OSBEP_pvec}).
An approach to determine $U_{\min}$ will be introduced later in Section \ref{subSec:OptSC_Umin}.
The constraint (\ref{Const_NEP_pvec}) indicates the maximum allowable NEP at the receiver where $\zeta(\mathbf{p})$ was defined in (\ref{NEP_CPSOFDM}). Note that when $\epsilon=0$ is chosen in (\ref{Const_NEP_pvec}), the equality holds if and only if the CPS precoding matrix $\mathbf{P}$ is unitary. The constraint (\ref{Const_FixEnergy_pvec}) is to fix the energy of the design variable $\mathbf{p}$.

To handle the problem (\ref{OptShapingVecProblem}), we adopt the idea in \cite{Sun2017_OPT_MM} and transform the order of objective function from quartic to quadratic by change of variables.
To this end, we first introduce a lifting matrix \cite{Sun2017_OPT_MM} $\mathbf{X}=\mathbf{p} \mathbf{p}^{H}$ to transform the order of ${\bar \sigma}_{x}^{2} ( \mathbf{p} ) $ from quartic to quadratic. 
Thereafter, the objective VIP function is rewritten as ${\bar \sigma}_{x}^{2}  \left( \mathbf{X} \right) = f(\mathbf{X}) - {\bar \mu}_{x}^{2} $ where $f(\mathbf{X}) = \frac{1}{N}  \sum_{n=0}^{N-1} \mathbb{E} \left\{ \left| x_{n} \right|^{4} \right\}$.
Since ${\bar \mu}_{x}$ is a constant, it can actually be dropped from the objective function. 
The function $f({\bf X})$ can be shown to be convex in ${\bf X}$ by the following reasoning.
Using (\ref{quartic_xn}) and letting $\mathbf{u}_{k,m,n}=\mathbf{C}_{kM}^{T}\mathbf{e}_{m,n}$, it can be shown that 
\begin{align}
\label{mxquadratic_xn}
\mathbb{E}   \left\{  \left|  x_{n} \right|^{4}  \right\}  &= \sigma_{\mathrm{d}}^{4} \sum_{\scriptstyle k\in \mathcal{K} \atop \scriptstyle m\in \mathcal{M}} \left( \mathbf{u}_{k,m,n}^{H}  \mathbf{X} \mathbf{u}_{k,m,n}^{}  \right)^2   \nonumber \\
+  2  E_{\mathrm{s}}^2 & \sum_{ \scriptstyle k\in \mathcal{K} \atop \scriptstyle  m\in \mathcal{M}} \sum_{ \substack{ k'\in \mathcal{K} \\  m'\in \mathcal{M} \\ (k',m') \neq (k,m) }} \left| \mathbf{u}_{k',m',n}^{H}  \mathbf{X} \mathbf{u}_{k,m,n}^{}  \right|^2,
\end{align}
where both $ ( \mathbf{u}_{k,m,n}^{H}  \mathbf{X} \mathbf{u}_{k,m,n}^{}  )^2 $ and $| \mathbf{u}_{k',m',n}^{H}  \mathbf{X} \mathbf{u}_{k,m,n}^{} |^2$ are convex functions of $\mathbf{X} \succeq \mathbf{0}_{S\times S}$. Thus, $f(\mathbf{X})$ is convex, since it is a nonnegative weighted sum of convex functions \cite{Boyd2004_TB}. 
We further proceed to rewrite $f({\bf X})$ in the quadratic form of $\mathrm{vec}({\bf X})$.
To do so, we define the matrix $\mathbf{U}_{k,m,n}=\mathbf{u}_{k,m,n}^{}\mathbf{u}_{k,m,n}^{H}$ and $\mathbf{\bar U}_{k,m,n}=\sum_{ \substack{ k'\in \mathcal{K} \\  m'\in \mathcal{M} \\ (k',m') \neq (k,m) }} \mathbf{U}_{k',m',n}$ and (\ref{mxquadratic_xn}) can be rewritten as
\begin{align}
\label{XUmxquadratic_xn}
\mathbb{E}  \left\{  \left|  x_{n} \right|^{4}  \right\}  &= \sigma_{\mathrm{d}}^{4} \sum_{\scriptstyle k\in \mathcal{K} \atop \scriptstyle m\in \mathcal{M}} \left[  \mathrm{tr} \left( \mathbf{X} \mathbf{U}_{k,m,n}^{}  \right) \right]^2   \nonumber \\
& +  2  E_{\mathrm{s}}^2  \sum_{ \scriptstyle k\in \mathcal{K} \atop \scriptstyle  m\in \mathcal{M}}   \mathrm{tr} \left( \mathbf{X} \mathbf{\bar U}_{k,m,n} \mathbf{X} \mathbf{U}_{k,m,n} \right) .
\end{align}
By using the formulae $\mathrm{tr} ( \mathbf{X} \mathbf{U}_{k,m,n}^{} ) = \mathrm{vec}(\mathbf{X})^{H} \mathrm{vec}( \mathbf{U}_{k,m,n})^{} $ and $ \mathrm{tr} ( \mathbf{X} \mathbf{\bar U}_{k,m,n} \mathbf{X} \mathbf{U}_{k,m,n} ) = \mathrm{vec}(\mathbf{X})^{H} (  \mathbf{U}_{k,m,n}^{T} \otimes \mathbf{\bar U}_{k,m,n}^{} )$ $\mathrm{vec}(\mathbf{X})^{} $ for (\ref{XUmxquadratic_xn}), we can finally obtain the objective function in matrix quadratic form \cite{Sun2017_OPT_MM}, i.e.,
\begin{eqnarray}
\label{MxQuadraticObjFun}
f ( \mathbf{X} )   =   \mathrm{vec} \left( \mathbf{X} \right)^{H} \mathbf{T}  \mathrm{vec} \left( \mathbf{X} \right),
\end{eqnarray}
where the $S^{2} \times S^{2}$ Hermitian positive definite matrix
\begin{align}
\label{TMx}
\mathbf{T} = \frac{1}{N}  \sum_{n=0}^{N-1}  \sum_{\scriptstyle k\in \mathcal{K} \atop \scriptstyle m\in \mathcal{M}} \Big[  \sigma_{\mathrm{d}}^{4} \enskip & \mathrm{vec} \left( \mathbf{U}_{k,m,n}^{} \right)^{} \mathrm{vec} \left( \mathbf{U}_{k,m,n}^{} \right)^{H}  \nonumber \\
&+ 2  E_{\mathrm{s}}^2 \left(  \mathbf{U}_{k,m,n}^{T} \otimes \mathbf{\bar U}_{k,m,n}^{}   \right)  \Big]
\end{align}
can be offline computed and stored before performing optimization procedures. Besides, it is easy to equivalently describe the constraints (\ref{Const_OSBEP_pvec})-(\ref{Const_FixEnergy_pvec}) in terms of the changed variable $\mathbf{X}$ \cite{Chi2017_TB}. 
For example, $\gamma_x({\bf p})$ defined in (\ref{OSBEP_CPSOFDM}) can be rewritten as $\mathrm{tr}(\mathbf{\Omega}\mathbf{X})$, and $\zeta({\bf p})$ defined in (\ref{NEP_CPSOFDM}) can be rewritten as 
$ \zeta(\mathbf{X}) =  \mathrm{tr}\left(   \left[  \left(  \mathbf{W}_{K}^{H} \otimes \mathbf{I}_{M} \right)  \mathbf{X}  \left(  \mathbf{W}_{K}^{H} \otimes \mathbf{I}_{M} \right)^{H}  \right]^{\circ -1 }   \right).$
As a result, the proposed optimization problem (\ref{OptShapingVecProblem}) is reformulated as
\begin{subequations}
\label{MxQuadraticMinProblem}
\begin{align}
\underset{\mathbf{X}}{\text{minimize}} \enskip 
&\ \label{ObjFun_VIP_Xmx}
\mathrm{vec} \left( \mathbf{X} \right)^{H} \mathbf{T}  \mathrm{vec} \left( \mathbf{X} \right)^{}  \\
\text{subject to} \enskip 
&\  \label{Const_OSBEP_Xmx}
\mathrm{tr} \left( \mathbf{\Omega}  \mathbf{X} \right)  \leq U \\
&\  \label{Const_NEP_Xmx}
\zeta \left( \mathbf{X} \right) \leq  (1+\epsilon) \frac{S^2}{\rho} \\
&\  \label{Const_FixEnergy_Xmx}
\mathrm{tr}(\mathbf{X}) = \rho \\
&\  \label{PSDMxConst}
\mathbf{X} \succeq \mathbf{0}_{S\times S} \\ 
&\   \label{RankOneConst}
\mathrm{rank}(\mathbf{X}) = 1,
\end{align}
\end{subequations}
in which only the rank-one constraint (\ref{RankOneConst}) is nonconvex.
In (\ref{Const_OSBEP_Xmx}), the matrix $\mathbf{\Omega}$ was defined right after (\ref{OSBEP_CPSOFDM}).

To solve the problem (\ref{MxQuadraticMinProblem}), we adopt the MM method \cite{Sun2017_OPT_MM} whose key idea is to convert a difficult problem into a series of simple problems with convergence guarantee \cite{Sun2017_OPT_MM}, \cite{Razaviyayn2013_OPT_MM}. The first majorization step of MM is to construct a surrogate function $g( \mathbf{X}^{} | \mathbf{X}^{(\ell)} )$ of (\ref{MxQuadraticObjFun}) defined as
\begin{align}
\label{UBofObjFun}
g \left( \mathbf{X}^{} | \mathbf{X}^{(\ell)} \right)  &=  
\mathrm{vec} \left( \mathbf{X} \right)^{H} \mathbf{D} \mathrm{vec} \left( \mathbf{X} \right)   \nonumber \\
& \enskip +
2 \Re \left\{ \mathrm{vec} \left( \mathbf{X} \right)^{H} \left[ \mathbf{T} - \mathbf{D} \right] \mathrm{vec} \left( \mathbf{X}^{(\ell)} \right)  \right\} \nonumber  \\
& \enskip +
\mathrm{vec} \left( \mathbf{X^{(\ell)}} \right)^{H} \left[ \mathbf{D} - \mathbf{T}  \right] \mathrm{vec} \left( \mathbf{X}^{(\ell)} \right) 
\end{align}
where $\mathbf{D}$ is a Hermitian matrix chosen such that $\mathbf{D} \succeq \mathbf{T}$.
Then, according to \cite[Eq. (26)]{Sun2017_OPT_MM},
it can be shown that $g(\mathbf{X} | \mathbf{X}^{(\ell)}) \geq f(\mathbf{X})$ for any $\mathbf{X}$ as long as $\mathbf{D} \succeq \mathbf{T}$ .
Note that $g( \mathbf{X}^{(\ell)} | \mathbf{X}^{(\ell)} )  = f ( \mathbf{X}^{(\ell)} )$. 
Now, we choose that $\mathbf{D} = \lambda_{\max} \left( \mathbf{T} \right) \mathbf{I}_{S^2} $ and simplify the surrogate function (\ref{UBofObjFun}) as  
\begin{align}
\label{SurrogateFun}
g \left( \mathbf{X}^{} | \mathbf{X}^{(\ell)} \right) &= 2 \Re  \left\{ \mathrm{vec} \left( \mathbf{X} \right)^{H} \mathbf{J}  \mathrm{vec} \left( \mathbf{X^{(\ell)}} \right)  \right\}  + c  \nonumber \\
&= 2 \Re  \left\{  \mathrm{tr} \left( \mathbf{E}^{(\ell)} \mathbf{X}^{} \right) \right\}  + c \nonumber \\
&= \mathrm{tr} \left( \mathbf{V}^{(\ell)} \mathbf{X} \right) + c ,
\end{align}
where $ \mathbf{J} =  \mathbf{T} - \lambda_{\max} \left( \mathbf{T} \right) \mathbf{I}_{S^2} $, $c$ is a constant term independent of $\mathbf{X} $, $\mathbf{E}^{(\ell)} = \mathrm{reshape} \left(  \mathbf{J}  \mathrm{vec} \left( 
\mathbf{X}^{(\ell)}\right) , S,S \right) $, and 
\begin{equation}
\mathbf{V}^{(\ell)}=\frac{1}{2} \left( {\mathbf{E}^{(\ell)}}^{} + {\mathbf{E}^{(\ell)}}^{H} \right).
\label{Vl}
\end{equation}
Based on (\ref{SurrogateFun}), the second minimization step of MM is to perform the following SDP with multiple iterations. 
\begin{subequations}
\label{SDPProblem}
\begin{align}
\underset{\mathbf{X}}{\text{minimize}} \enskip
&\ \label{SDPObjFun_VIP_Xmx}
\mathrm{tr} \left( \mathbf{V}^{(\ell)}\mathbf{X} \right)  \\
\text{subject to} \enskip
&\  \label{SDPConst_OSBEP_Xmx}
\mathrm{tr} \left( \mathbf{\Omega}  \mathbf{X} \right)  \leq U \\
&\  \label{SDPConst_NEP_Xmx}
\zeta \left( \mathbf{X} \right) \leq  (1+\epsilon) \frac{S^2}{\rho} \\
&\  \label{SDPConst_FixEnergy_Xmx}
\mathrm{tr}(\mathbf{X}) = \rho \\
&\  \label{SDPPSDMxConst}
\mathbf{X} \succeq \mathbf{0}_{S\times S}.
\end{align}
\end{subequations}
Note that although the rank-one constraint has been relaxed in (\ref{SDPProblem}), from empirical results we observe that $\mathrm{rank}(\mathbf{X}^{(\ell)}) = 1$ always holds. 
The optimal solution to (\ref{OptShapingVecProblem}) can therefore be obtained by finding $\mathbf{p}_{\mathrm{opt}}^{}$ that satisfies
$\mathbf{X}_{\mathrm{opt}}=\mathbf{p}_{\mathrm{opt}}^{}\mathbf{p}_{\mathrm{opt}}^{H} $.




\begin{algorithm}[t]
\caption{Proposed MM-CI-based optimization procedure for optimal prototype shaping vector design of CPS-OFDM.}
\begin{algorithmic}[1]
\renewcommand{\algorithmicrequire}{\textbf{Input:}}
\renewcommand{\algorithmicensure}{\textbf{Output:}}
\REQUIRE $S$, $\mathbf{\Omega}$ (\ref{OSBEP_CPSOFDM}), $\mathbf{T}$ (\ref{TMx}), $\rho$, $\epsilon$, $w$, $\beta$, $\varepsilon_{\mathrm{CI}}$, $\varepsilon_{\mathrm{MM}}$.
\ENSURE Optimal $S\times 1$ prototype shaping vector $\mathbf{p}_{\mathrm{opt}} $.
\\ \textit{CI Process}:
\STATE Initialize $\mathbf{B}^{(0)}=\mathbf{0}_{S\times S}$, set $\varphi=0$.
\REPEAT
\STATE Solve Problem (\ref{minOSBEPProblem}) to get the optimal solution $\mathbf{Y}^{(\varphi)}$.
\STATE Calculate the OSBEP $U^{(\varphi)}=\mathrm{tr} \left( \mathbf{\Omega}  \mathbf{Y}^{(\varphi)} \right)$.
\STATE Obtain $\mathbf{B}^{(\varphi+1)}$ according to (\ref{BMx}).
\STATE $\varphi \leftarrow \varphi+1$
\UNTIL convergence, i.e., $| U^{(\varphi+1)} - U^{(\varphi)} | \leq \varepsilon_{\mathrm{CI}} $.
\STATE Obtain $\mathbf{Y}_{\min} = \mathbf{Y}^{(\varphi)}$ and $U_{\min} = U^{(\varphi)}$.
\\ \textit{MM Process}:
\STATE Initialize $\mathbf{X}^{(0)}=\mathbf{Y}_{\min}$ and $\mathbf{V}^{(0)}$ according to (\ref{Vl}).
\STATE Set $U = \beta U_{\min}$ and $\ell=0$. 
\REPEAT
\STATE Solve Problem (\ref{SDPProblem}) to get the optimal solution $\mathbf{X}^{(\ell+1)}$.
\STATE Calculate the objective value $g^{(\ell)}=\mathrm{tr} \left( \mathbf{V}^{(\ell)}\mathbf{X}^{(\ell+1)} \right) $.
\STATE Obtain $\mathbf{V}^{(\ell+1)}$ according to (\ref{Vl}).
\STATE $\ell \leftarrow \ell+1$
\UNTIL convergence, i.e., $| g^{(\ell+1)} - g^{(\ell)} | \leq  \varepsilon_{\mathrm{MM}} $.
\STATE Obtain $\mathbf{p}_{\mathrm{opt}}=\mathbf{p}^{(\ell+1)}$ from $\mathbf{X}^{(\ell+1)} = {\mathbf{p}^{(\ell+1)}}^{}{\mathbf{p}^{(\ell+1)}}^{H} $.
\end{algorithmic} 
\end{algorithm}

\subsection{Choices of OSBEP Upper Bound and Initial Point in (\ref{SDPProblem})}
\label{subSec:OptSC_Umin}
Given the iterative MM algorithm presented earlier, it remains to find an initial point $\mathbf{X}^{(0)}$. 
While it is suggested in \cite{Zhao2017_OPT_MM} that some random initialization might work, in this subsection we consider this together with the problem of choosing the OSBEP upper bound $U$ in (\ref{SDPConst_OSBEP_Xmx}).
In this study, we choose $U = \beta U_{\min}$, where $U_{\min} >0$ is the minimum value of the OSBEP and $\beta \geq 1$ is a factor we can choose. To find $U_{\min} $, the following optimization problem is first solved under the convex-iteration (CI) algorithmic framework \cite{Dattorro2017_TB} 
\begin{subequations}
\label{minOSBEPProblem}
\begin{align}
\underset{\mathbf{Y}}{\text{minimize}} \enskip
&\ \label{minOSBEPObjFun_OSBEP_Xmx}
w\cdot \mathrm{tr} \left( \mathbf{Y} \mathbf{B}^{(\varphi)} \right)  + \mathrm{tr} \left( \mathbf{\Omega}  \mathbf{Y} \right)  \\
\text{subject to} \enskip
&\  \label{minOSBEPConst_NEP_Xmx}
\zeta \left( \mathbf{Y} \right) \leq  (1+\epsilon) \frac{S^2}{\rho} \\
&\  \label{minOSBEPConst_FixEnergy_Xmx}
\mathrm{tr}(\mathbf{Y}) = \rho \\
&\  \label{minOSBEPPSDMxConst}
\mathbf{Y} \succeq \mathbf{0}_{S\times S},
\end{align}
\end{subequations}
where $w>0$ is an empirical weighting factor and the matrix
\begin{eqnarray}
\label{BMx}
\mathbf{B}^{(\varphi)} = [\mathbf{\tilde U}]_{\{1,2,\cdots,S-1 \}}^{} [\mathbf{\tilde U}]_{\{1,2,\cdots,S-1 \}}^{H} 
\end{eqnarray}
is derived from the singular value decomposition (SVD) of $\mathbf{Y} = \mathbf{\tilde U} \mathbf{\tilde \Sigma} \mathbf{\tilde V}$ at the $(\varphi-1)$th iteration. After the algorithm converges, we obtain $U_{\min}=  \mathrm{tr} \left( \mathbf{\Omega}  \mathbf{Y}_{\min} \right) $, where the optimal rank-one point $\mathbf{Y}_{\min}=\mathbf{p}_{\min}^{} \mathbf{p}_{\min}^{H}$ also serves as the initial point $\mathbf{X}^{(0)}$ locating in the feasible set of the problem (\ref{SDPProblem}). Consequently, the proposed MM-CI-based optimization procedure can be summarized in Algorithm 1. To solve the problems (\ref{SDPProblem}) and (\ref{minOSBEPProblem}), we adopt \texttt{CVX}, a package for specifying and solving convex programs \cite{CVXtool}. The tolerant precision for the stopping criteria of the CI process and the MM process are denoted by $\varepsilon_{\mathrm{CI}}>0$ and $\varepsilon_{\mathrm{MM}}>0$, respectively.

\begin{figure*}[t]
\centering \centerline{
\includegraphics[width=1.0\textwidth,clip]{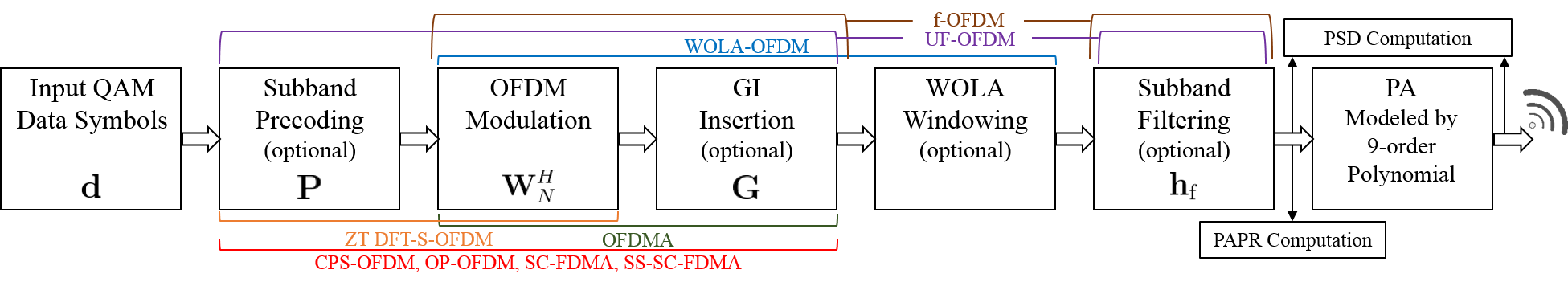}}
\caption{A generalized simulation flowchart of different waveform generation at the transmitter.}
\label{TxFlowchart}
\end{figure*}

\section{Performance Evaluations in 5G NR}
\label{Sec:SimResult}
In this section, computer simulations are conducted to demonstrate the advantages of the proposed CPS-OFDM over other existing waveform candidates in 5G NR. A generalized simulation flowchart of different waveform generation at the transmitter is depicted in Fig. \ref{TxFlowchart}.


\subsection{Simulation Parameters, Assumptions, and Cases}
\label{subSec:SimSetup}
In our simulator, the system parameters are practically chosen according to the agreements in 3GPP standardization meetings for 5G NR waveform evaluation \cite[Annex A.1.1]{3GPPTR38802}. The carrier frequency is at $4$ GHz. The sampling rate is $15.36$ MHz. The FFT size $N$ is $1024$ with $15$ kHz subcarrier spacing. The GI length $G$ is $72$. Uncoded 16QAM is utilized with ${E_{\mathrm{s}}}=1$. The corresponding $\sigma_{\mathrm{d}}^{4}$ can be derived as 1.32. The bit power versus noise variance is denoted by ${E_{\mathrm{b}}}/N_{0}$. The bandwidth assigned to each UE $\mathrm{BW}_{\mathrm{UE}}$ is $720$ kHz, namely, $S=48$. The guard band between two users is $60$ kHz. The ninth-order polynomial approximation specified in \cite{3GPP166004_3GPPTDOC} is used to model PA nonlinearity with the phase compensation of $76.3$ degrees \cite{Huawei166093_3GPPTDOC}. A challenging IBO value of $3$ dB is selected to investigate the achievable performance, while the spectral regrowth fulfills the spectrum emission mask (SEM) defined in \cite[Table 6.6.2.1.1-1]{3GPPTS36101} with the maximum UE transmit power of $22$ dBm. The configuration of single transmit antenna and single receive antenna (1T1R) is set. The tapped delay line (TDL)-C channel model is used in terms of the delay spread scaled by $300$ ns with $3$ km/h mobility \cite[Table 7.7.2-3]{3GPPTR38901}. The MMSE-FDE (\ref{MMSEFDEMx}) is adopted at the receiver under the assumption of perfect channel estimation.


All uplink cases recorded in \cite[Sec. 7.1.1]{3GPPTR38802}, namely, Case 1b, Case 3, and Case 4, are taken into consideration (note that Cases 1a and 2 in \cite[Sec. 7.1.1]{3GPPTR38802} are downlink cases and are out of the scope of this article). 
In Case 1b, there is only a target user allocated on the subcarriers indexed by $ \mathcal{I}_{}=\left\{ 212,213,\cdots, 259 \right\}$. In Case 3, one target user and two asynchronous interfering users are assigned to the subcarriers indexed by $ \mathcal{I}_{}=\left\{ 488,489,\cdots, 535 \right\}$, $\mathcal{I}_{1}=\left\{ 540,541,\cdots, 587 \right\}$, and $\mathcal{I}_{2}=\left\{ 436,437,\cdots, 483 \right\}$, respectively. The timing offset (TO) of the target user is assumed to be perfectly estimated and compensated at the receiver. The relative TOs of the interfering users in terms of $128$ delayed samples incur multiuser interference. In Case 4, there are one target user and two interfering users in synchronism but in different numerology. The target user complies with the same setting as in Case 3 in terms of the basic $15$ kHz subcarrier spacing. The two interfering users adopting $30$ kHz subcarrier spacing are assigned to the subcarriers indexed by $\mathcal{I}_{1}= \left\{ 270,271,\cdots, 293 \right\}$ and $\mathcal{I}_{2}=\left\{ 218,219,\cdots, 241 \right\}$, whose FFT size and GI length are $512$ and $36$, respectively.

\begin{figure}[t]
\centering
   \subfigure[for the comparisons with low OSBE waveforms]{\includegraphics[width=0.5\textwidth,clip]{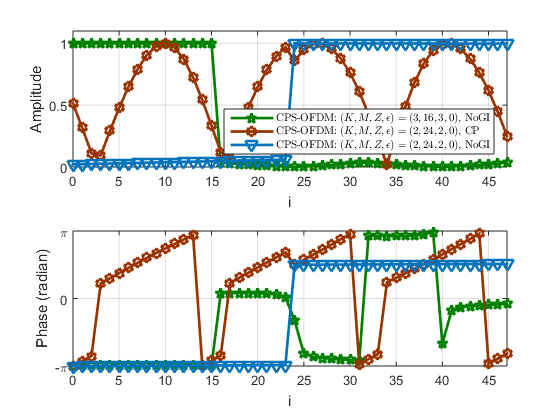}} 
   \subfigure[for the comparisons with low PAPR waveforms]{\includegraphics[width=0.5\textwidth,clip]{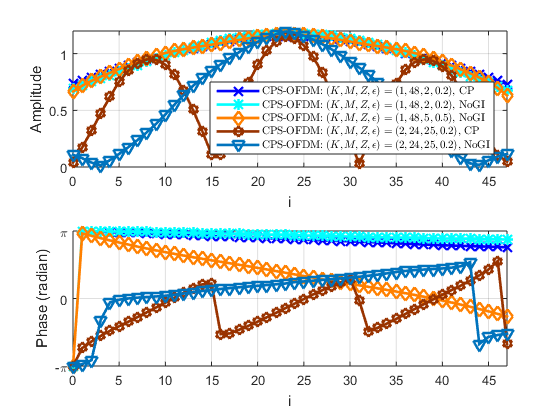}}
   \caption{Illustrations of the optimized prototype shaping vectors by Algorithm 1 with respect to different parameter settings of CPS-OFDM in Case 1b.}
   \label{Result_Case1b_pvec}
\end{figure}

\subsection{Settings of Existing Waveforms for Comparisons}
Two 4G legacy waveforms, OFDMA ($\mathbf{P}=\mathbf{I}_{S}$, $\mathbf{G}=\mathbf{G}_{\mathrm{cp}}$) and SC-FDMA ($\mathbf{P}=\mathbf{W}_{S}$, $\mathbf{G}=\mathbf{G}_{\mathrm{cp}}$), without the utilization of zero symbols ($Z=0$), serve as the performance benchmark. Other waveforms for comparisons are set as below. 
Note that all waveforms can be expressed as a special case of the generalized form depicted in Fig. \ref{TxFlowchart}.
 
\subsubsection{WOLA-OFDM ($\mathbf{P}=\mathbf{I}_{S}$, $\mathbf{G}=\mathbf{G}_{\mathrm{cp}}$, $Z=0$)} 
The realization of WOLA processing is based on \cite[Fig. 2-13]{Qualcomm162199_3GPPTDOC}, in which the raised cosine (RC) window length is chosen to be $2G$, where $G$ is the GI length defined before (\ref{TxSignal}), with the roll-off factor $\alpha=1$ \cite{YliKaakinen2017_WD}, \cite{Nokia165013_3GPPTDOC}. 

\subsubsection{UF-OFDM ($\mathbf{P}=\mathbf{P}_{\mathrm{preeq}}$, $\mathbf{G}=\mathbf{G}_{\mathrm{zp}}$, $Z=0$)}
The transceiver model is provided in \cite{Nokia165014_3GPPTDOC}. The order of the Dolph-Chebyshev filter $\mathbf{h}_{\mathrm{f}}$ is $L_{\mathrm{f}}=G$ with the attenuation factor of $75$ dB \cite{Nokia165013_3GPPTDOC}. The pre-equalization of the filter response is adopted, i.e., $\mathbf{P}_{\mathrm{preeq}}= \mathrm{diag} ( [ \mathbf{W}_{N} [ \mathbf{h}_{\mathrm{f}}^{T} \enskip \mathbf{0}_{(N-L_{\mathrm{f}}-1)\times 1}^{T} ]^{T} ]_{\mathcal{I}}^{\circ -1} )$.

\subsubsection{f-OFDM ($\mathbf{P}=\mathbf{I}_{S}$, $\mathbf{G}=\mathbf{G}_{\mathrm{cp}}$, $Z=0$)}
The detailed filter design procedure is given in \cite{Huawei165425_3GPPTDOC}. In brief, the filter $\mathbf{h}_{\mathrm{f}}$ is based on the RC-windowed tone offset sinc function. Two filter orders $L_{\mathrm{f}}=N/2$ and $L_{\mathrm{f}}=G$ (i.e., half-block and GI lengths) are of interest with the tone offset being $N_{\mathrm{TO}}=2.5$ and $N_{\mathrm{TO}}=5$, respectively.

\subsubsection{OP-OFDM ($ \mathbf{P} =[ \mathbf{0}_{S\times Z} \enskip \mathbf{P}_{\mathrm{o}} ]$, $\mathbf{G}=\mathbf{G}_{\mathrm{cp}}$, $Z=2$)} 
Orthogonal precoding for OFDM (OP-OFDM) is proposed in \cite{Ma2011_WD_POFDM} to suppress the spectral leakage power without NEP. The $S\times D$ precoding matrix $\mathbf{P}_{\mathrm{o}}$ is determined by \cite[Eq. (10)]{Ma2011_WD_POFDM}. The input data symbols are corresponding to $\mathcal{D}=\{ Z,Z+1,\cdots,S-1 \}$.

\subsubsection{ZT DFT-S-OFDM ($\mathbf{P}=\mathbf{W}_{S}$, $\mathbf{G}=\mathbf{I}_{N}$, $Z=1+\left \lceil SG/N \right \rceil $)}
The internal GI is realized by setting $\mathcal{D}=\{ 1,2,\cdots,S-Z \}$ for the input data vector $\mathbf{d}$, where $Z=Z_{\mathrm{head}}+Z_{\mathrm{tail}}$, $Z_{\mathrm{head}}=1$, $Z_{\mathrm{tail}}=\left \lceil SG/N \right \rceil $ \cite{Berardinelli2013_WD_DFTSOFDM}.

\subsubsection{SS-SC-FDMA ($\mathbf{P} =[ \mathbf{0}_{S\times Z/2} \enskip \mathbf{P}_{\mathrm{ss}} \enskip \mathbf{0}_{S\times Z/2} ]$, $\mathbf{G}=\mathbf{G}_{\mathrm{cp}}$, $Z=S/2$)}
This scheme is to reduce the PAPR at the cost of excess bandwidth and NEP. We consider the extreme case that only $D=S/2$ data symbols are transmitted within the UE bandwidth. The $S\times D$ precoding matrix is obtained by $\mathbf{P}_{\mathrm{ss}}= \mathrm{diag} \left( \mathbf{p}_{\mathrm{RRC}} \right) ( \mathbf{1}_{2} \otimes \mathbf{I}_{D}) \mathbf{W}_{D}$, where $\mathbf{p}_{\mathrm{RRC}}$ is the $S\times 1$ root raised cosine (RRC) shaping vector with the roll-off factor $\alpha=1$, $\mathcal{D}=\{ S/4,S/4+1,\cdots,S/4+S/2-1 \}$ \cite{Kawamura2006_WD_DFTSOFDM}.

In the first phase of 5G NR standardization, it is widely accepted that windowing and filtering techniques at the transmitter are transparent to the receiver \cite{3GPPTR38912}. Hence, the receiver model shown in Fig. \ref{TxRx_PrecOFDM} can be directly applied to the aforementioned waveform schemes. For WOLA-OFDM, the WOLA processing at the receiver \cite[Fig. 2-15]{Qualcomm162199_3GPPTDOC} is omitted. For UF-OFDM and f-OFDM, the received frequency-domain signal (\ref{procRxSignal}) is rewritten as $ \mathbf{r} =  \mathbf{H}_{\mathrm{c}} \mathbf{P}\mathbf{d} + \mathbf{v}$,
where $\mathbf{H}_{\mathrm{c}} = \mathbf{H} \cdot \mathrm{diag}( [ \mathbf{W}_{N} [ \mathbf{h}_{\mathrm{f}}^{T} \enskip \mathbf{0}_{(N-L_{\mathrm{f}}-1)\times 1}^{T} ]^{T} ]_{\mathcal{I}} )$ is the composite channel frequency response.




\begin{figure}[t]
\centering \centerline{
\includegraphics[width=0.5\textwidth,clip]{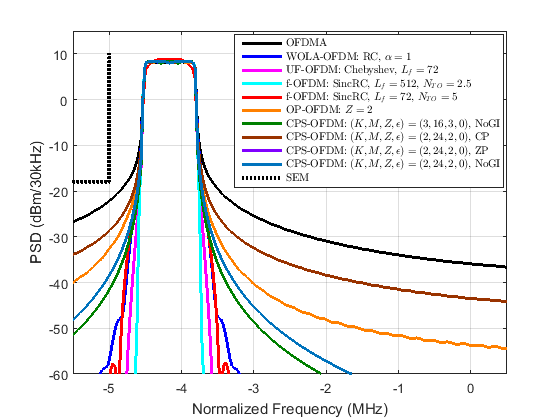}} 
\caption{Simulated PSD results of the waveforms claiming low OSBE in the absence of PA, where WOLA-OFDM, UF-OFDM, and f-OFDM ideally have extremely low OSBE at the cost of increased PAPR and induced IBI.}
\label{Result_Case1b_PSD_K2}

\centering \centerline{
\includegraphics[width=0.5\textwidth,clip]{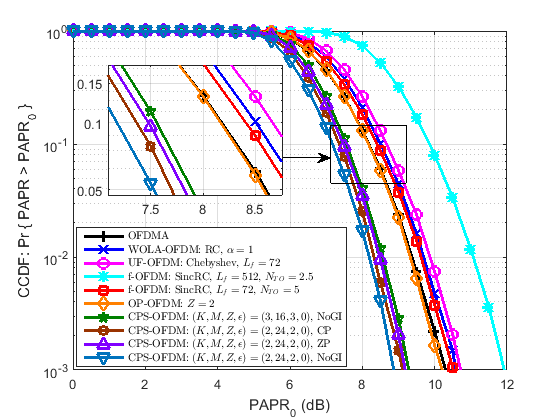}} 
\caption{PAPR performance results of the waveforms claiming low OSBE, where CPS-OFDM yields much lower PAPR and so ensures better PA efficiency as compared to the others.}
\label{Result_Case1b_PAPR_K2}

\end{figure}

\begin{figure}[t]
\centering \centerline{
\includegraphics[width=0.5\textwidth,clip]{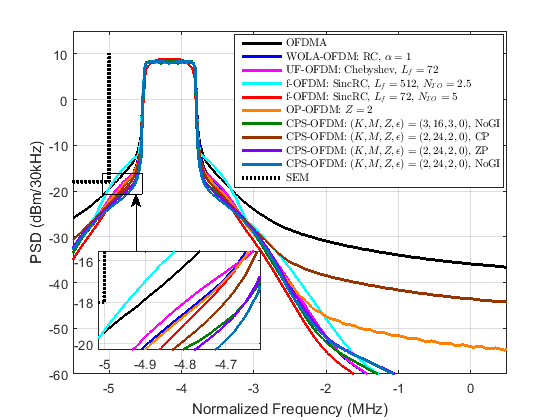}} 
\caption{Simulated PSD results of the waveforms claiming low OSBE in the presence of PA with the IBO of $3$ dB, where CPS-OFDM in practice leads to the lowest amount of OSBE in adjacent bands because of its low PAPR.}
\label{Result_Case1b_PSDPA_K2}

\centering \centerline{
\includegraphics[width=0.5\textwidth,clip]{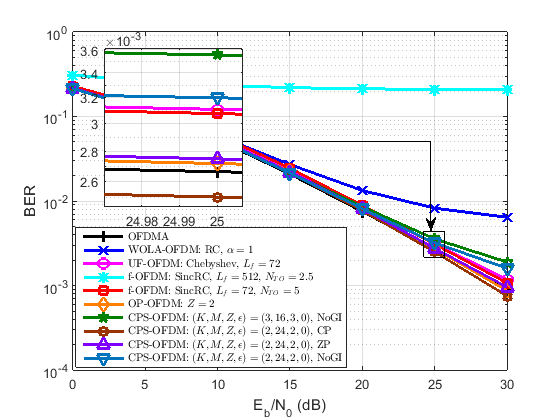}} 
\caption{Single user detection performance results in terms of uncoded 16QAM, IBO of $3$ dB, and TDL-C-300 channel, where CPS-OFDM with CP possesses the best signal reliability at the receiver.}
\label{Result_Case1b_BER_MMSEPA_K2}

\end{figure}

\begin{figure}[t]
\centering \centerline{
\includegraphics[width=0.5\textwidth,clip]{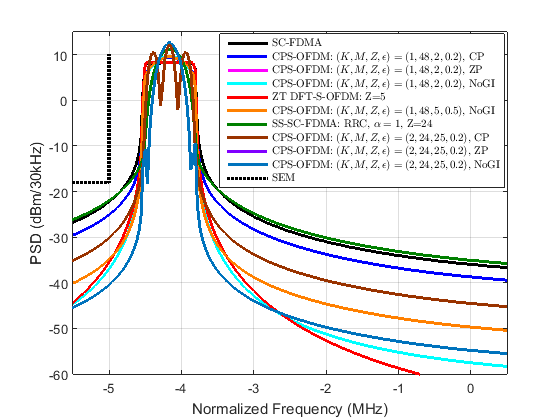}} 
\caption{Simulated PSD results of the waveforms claiming low PAPR in the absence of PA, where SC-FDMA and SS-SC-FDMA do not address the OSBE issue and the controllable passband fluctuation of CPS-OFDM can be seen.}
\label{Result_Case1b_PSD_K1}

\centering \centerline{
\includegraphics[width=0.5\textwidth,clip]{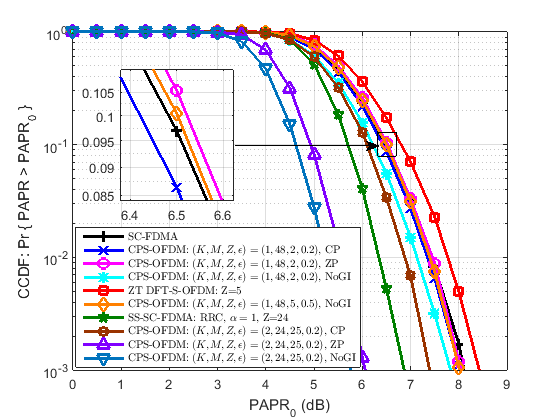}} 
\caption{PAPR performance results of the waveforms claiming low PAPR, where CPS-OFDM can further reduce the PAPR by allowing some NEP ($\epsilon >0$) for the optimal prototype shaping vector design.}
\label{Result_Case1b_PAPR_K1}
\end{figure}

\begin{figure}[t]
\centering \centerline{
\includegraphics[width=0.5\textwidth,clip]{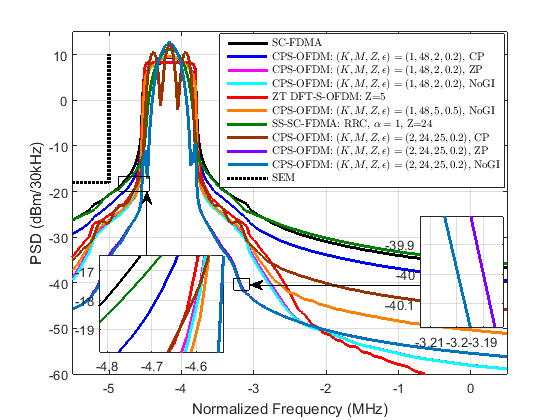}} 
\caption{Simulated PSD results of the waveforms claiming low PAPR in the presence of PA with the IBO of $3$ dB, where CPS-OFDM benefited from the design flexibility can handle the issues of OSBE, PAPR, and NEP adaptively.}
\label{Result_Case1b_PSDPA_K1}

\centering \centerline{
\includegraphics[width=0.5\textwidth,clip]{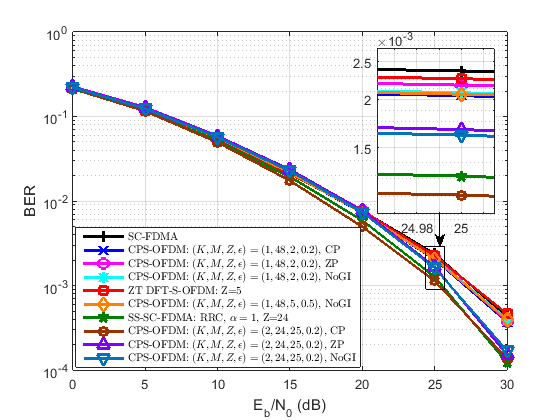}} 
\caption{Single user detection performance results in terms of uncoded 16QAM, IBO of $3$ dB, and TDL-C-300 channel, where CPS-OFDM with $|\mathcal{K}|<K$ can further improve the BER thanks to the frequency diversity.}
\label{Result_Case1b_BER_MMSEPA_K1}
\end{figure}

\subsection{Settings of The Proposed CPS-OFDM Waveform}
CPS-OFDM possesses a flexible DFT-based precoder (\ref{CPSPrec_direct}) parameterized by $K$, $M$, and $\mathbf{p}$. As the current evaluation cases belong to sub-6-GHz narrowband transmission, we consider 
\begin{itemize}
\item
$(K,M,Z,\epsilon)=(2,S/2,2,0)$ and $(3,S/3,3,0)$ for the comparisons with OFDMA, WOLA-OFDM, UF-OFDM, f-OFDM, and OP-OFDM without NEP, $\mathcal{K}=\mathcal{Z}_{K}$, $\mathcal{M}=\{ 1,2,\cdots,M-1\}$.
\item
$(K,M,Z,\epsilon)=(1,S,2,0.2)$ for the comparison with SC-FDMA, $\mathcal{K}=\mathcal{Z}_{1}$, $\mathcal{M}=\{ 1,2,\cdots,M-Z\}$.
\item
$(K,M,Z,\epsilon)=(1,S,1+\left \lceil SG/N \right \rceil ,0.5)$ and $\mathbf{G}=\mathbf{I}_{N}$ for the comparison with ZT DFT-S-OFDM, $\mathcal{K}=\mathcal{Z}_{1}$, $\mathcal{M}=\{ 1,2,\cdots,M-Z\}$.
\item
$(K,M,Z,\epsilon)=(2,S/2,M+1,0.2)$ for the comparison with SS-SC-FDMA, $\mathcal{K}=\{ 0 \}$, $\mathcal{M}=\{ 1,2,\cdots,M-1\}$.
\end{itemize}
Given these parameters with different GI types, the prototype shaping vector $\mathbf{p}$ is then determined by the proposed MM-CI-based optimization procedure in Algorithm 1. The OSBEP upper bound factor $\beta$ is $10$. The weighting factor $w$ is $1000$. The energy of $\mathbf{p}$ is $\rho=M$. The maximum number of CI and MM iterations are $10^{4}$ and $10^{5}$ with $\varepsilon_{\mathrm{CI}}=10^{-8}$ and $\varepsilon_{\mathrm{MM}}=10^{-10}$, respectively. The number of discretization samples per subcarrier spacing is $10$. For Case 1b, the OSB range $\mathcal{F}_{\mathrm{OSB}}$ corresponds to the subcarriers indexed by $\left\{ 0,\cdots,207,264,\cdots,1023 \right\}$. The optimized prototype shaping vectors with respect to different parameter settings are illustrated in Fig. \ref{Result_Case1b_pvec}. Note that CPS-OFDM with ZP and without GI (denoted by NoGI) reach the same result because of $G'=0$ in (\ref{PhiMx_entry})-(\ref{OSBEP_CPSOFDM}). For Case 3, the target user, the first interfering user, and the second interfering user treat the subcarriers indexed by $\left\{ 0,\cdots, 483, 540,\cdots,1023 \right\}$, $\left\{ 0,\cdots, 535, 592,\cdots,1023 \right\}$, and $\left\{ 0,\cdots, 431, 488,\cdots,1023 \right\}$ as their OSB ranges, respectively. For Case 4, the target user, the first interfering user, and the second interfering user treat the subcarriers indexed by $\left\{ 0,\cdots, 483, 540,\cdots,1023 \right\}$, $\left\{ 0,\cdots, 267, 296,\cdots,511 \right\}$, and $\left\{ 0,\cdots, 215, 244,\cdots,511 \right\}$ as their OSB ranges, respectively. Note that using precoding techniques for lowering OSBE generally demand some input zero symbols (i.e., $Z>0$). In return of it, the number of required guard subcarriers between two frequency domain adjacent users for lack of orthogonality can be significantly decreased if the spectral regrowth is limited. 

\begin{figure}[t]
\centering \centerline{
\includegraphics[width=0.5\textwidth,clip]{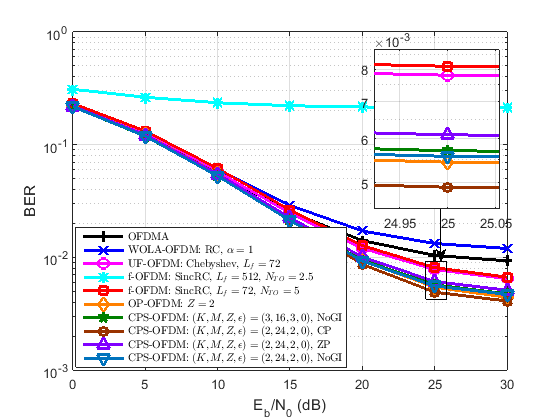}} 
\caption{Target user BER performance comparison of the waveforms claiming low OSBE in the asynchronous multiuser uplink case with IBO of $3$ dB, where CPS-OFDM with CP is much more robust to the TO than the others.}
\label{Result_Case3_BER_MMSEPA_K2}

\centering \centerline{
\includegraphics[width=0.5\textwidth,clip]{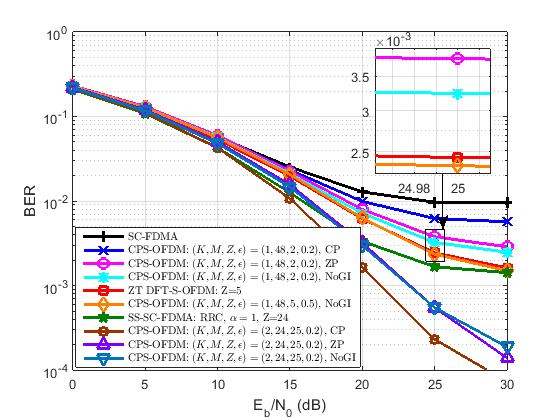}} 
\caption{Target user BER performance comparison of the waveforms claiming low PAPR in the asynchronous multiuser uplink case with IBO of $3$ dB, where CPS-OFDM can outperform the others by selecting proper parameters.}
\label{Result_Case3_BER_MMSEPA_K1}
\end{figure}

\begin{figure}[t]
\centering \centerline{
\includegraphics[width=0.5\textwidth,clip]{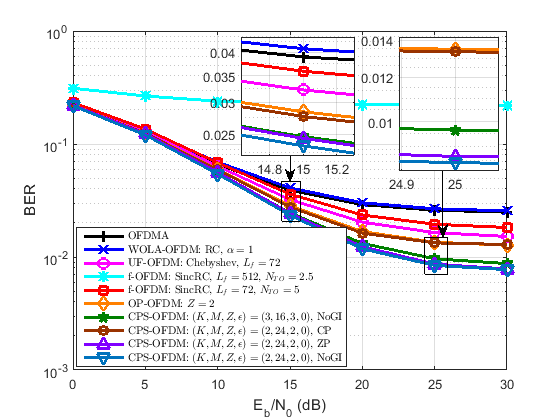}} 
\caption{Target user BER performance comparison of the waveforms claiming low OSBE in the mixed numerology multiuser uplink case with IBO of $3$ dB, where the superiority of CPS-OFDM over the others can be found.}
\label{Result_Case4_BER_MMSEPA_K2}

\centering \centerline{
\includegraphics[width=0.5\textwidth,clip]{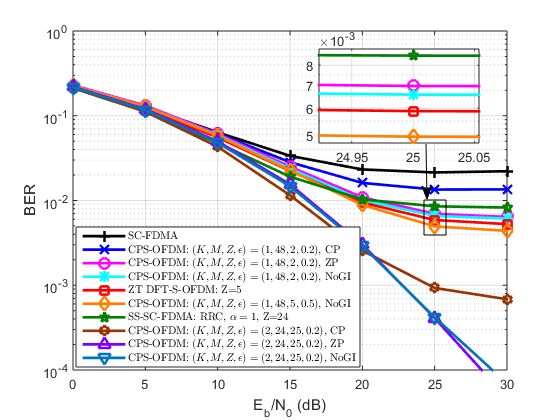}} 
\caption{Target user BER performance comparison of the waveforms claiming low PAPR in the mixed numerology multiuser uplink case with IBO of $3$ dB, where CPS-OFDM offers better detection reliability compared to the others.}
\label{Result_Case4_BER_MMSEPA_K1}
\end{figure} 
 
\subsection{Case 1b: Interference-Free Single User Performance}
Figure \ref{Result_Case1b_PSD_K2} depicts the simulated PSD results of CPS-OFDM and other waveforms that claim low OSBE, including WOLA-OFDM, UF-OFDM, f-OFDM, and OP-OFDM, in the absence of PA. 
We observe that CPS-OFDM with ZP and NoGI have the same outcome. 
The corresponding PAPR performances are reported in Fig. \ref{Result_Case1b_PAPR_K2}.
Here, the PAPR performance is assessed by plotting empirical complementary cumulative distribution function (CCDF) curves. Specifically, the PAPR is computed by \cite{Myung2008_TB}
\begin{align}
\label{PAPRDef}
\mathrm{PAPR}[b] =\frac{ \left \| \mathbf{\tilde x}[b]  \right \|_{\infty}^{2} }{  {N}_{\mathrm{ob}}^{-1}  \left \| \mathbf{\tilde x}[b]  \right \|_{2}^{2} },
\end{align}
where the ${N}_{\mathrm{ob}}\times 1$ vector $\mathbf{\tilde x}[b] $ is the $b$th transmitted block signal in terms of $J$-times oversampling (e.g., ${N}_{\mathrm{ob}}=J(N+G+L_{\mathrm{f}})$ for f-OFDM) (here, $J=4$ \cite{Rahmatallah2013_WDIS_PAE_PAPR}). 
The CCDF is defined as $1-\mathrm{Pr}\left\{ \mathrm{PAPR} \leq \mathrm{PAPR}_{0} \right\}$, where $\mathrm{Pr}\left\{ \mathrm{PAPR} \leq \mathrm{PAPR}_{0} \right\}$ indicates the probability of the PAPR that does not exceed a given threshold $\mathrm{PAPR}_{0}$. Observing the results shown in Fig. \ref{Result_Case1b_PAPR_K2}, CPS-OFDM yields much lower PAPR and so ensures better PA efficiency compared to the others. Notice that NoGI is more friendly to the PA than ZP by avoiding on-off switch. 
When the practical PA nonlinearity is considered, the corresponding effects of spectral regrowth can be observed in Fig. \ref{Result_Case1b_PSDPA_K2}. 
CPS-OFDM leads to the lowest amount of OSBE in adjacent bands. This may be due to its precoder design that addresses PAPR reduction at the same time. 
The required guard band, denoted $\Delta$, to warrant the OSBEP less than $-18$ dBm per $30$ kHz measurement bandwidth can be diminished so as to earn spectral efficiency. As $K$ increases, the OSBEP decreases in the ideal case (see Fig. \ref{Result_Case1b_PSD_K2}), but in the presence of PA nonlinearity, it actually depends on the spectral regrowth related to the PAPR increase and the choice of IBO. On the other hand, the expected spectral containment properties of WOLA-OFDM, UF-OFDM, and f-OFDM severely deteriorate due to their high PAPR drawbacks. In Fig. \ref{Result_Case1b_BER_MMSEPA_K2}, the bit error rate (BER) results of applying these waveforms to point-to-point communication are provided. In the presence of GI, CPS-OFDM possesses better detection performance than those of WOLA-OFDM, UF-OFDM, and f-OFDM suffering from the induced IBI and the significant PA nonlinear distortion. Compared to OP-OFDM, CPS-OFDM enjoys lower OSBE in adjacent bands, PAPR, BER, and implementation complexity.

Now, we turn our attention to existing waveforms that claim low PAPR, including SC-FDMA, ZT DFT-S-OFDM, SS-SC-FDMA, and compare that with the proposed CPS-OFDM. 
As a generalized form, CPS-OFDM can further reduce the PAPR by allowing some NEP (i.e., $\epsilon > 0$) and suppress the OSBE at the same time. Recall that in Fig. \ref{Result_Case1b_pvec}(b), the PSD passband fluctuation of CPS-OFDM owing to $\epsilon >0$ can be seen in Fig. \ref{Result_Case1b_PSD_K1}. The PAPR performance curves and the spectral regrowth results are displayed in Fig. \ref{Result_Case1b_PAPR_K1} and Fig. \ref{Result_Case1b_PSDPA_K1}, respectively. For CPS-OFDM, choosing either ZP or NoGI is generally easier to meet both low OSBE and low PAPR requirements than CP. The BER results are given in Fig. \ref{Result_Case1b_BER_MMSEPA_K1}. Benefited from the design flexibility and the optimized prototype shaping vector, CPS-OFDM is capable of achieving much lower OSBE, PAPR, and BER than those of SC-FDMA, ZT DFTS-OFDM and SS-SC-FDMA. For lack of additional DoF, the PA efficiency of ZT DFT-S-OFDM becomes worse as the number of input zeros increases \cite{Berardinelli2013_WD_DFTSOFDM}.

\begin{figure}[t]
\centering \centerline{
\includegraphics[width=0.5\textwidth,clip]{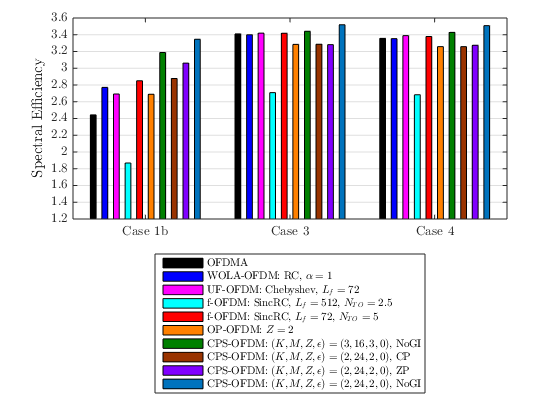}} 
\caption{Spectral efficiency comparisons of the waveforms claiming low OSBE in the interference-free single user case, the asynchronous multiuser case, and the mixed numerology multiuser case at $E_{\mathrm{b}}/N_{0}=25$ dB.}
\label{Result_SE_K2}
\end{figure} 

\subsection{Case 3: Asynchronous Multiuser Uplink Performance}
Recall that in Case 3, multiple asynchronous users, particularly with timing offsets (TO), are considered in performance evaluation.
The detection performance of target user applying different waveforms, in the presence of asynchronous interfering users, is provided in this subsection. 
Observing the BER results in Fig. \ref{Result_Case3_BER_MMSEPA_K2}, CPS-OFDM is more robust to TO than WOLA-OFDM, UF-OFDM, f-OFDM, and OFDMA. Given $Z=2$ and CP utilization, CPS-OFDM can achieve lower BER in terms of lower complexity than those of OP-OFDM. 
Without the WOLA processing at the receiver \cite[Fig. 2-15]{Qualcomm162199_3GPPTDOC}, it is very difficult for WOLA-OFDM to suppress the multiuser interference due to asynchronism. On the other hand, CPS-OFDM also outperforms SC-FDMA, ZT DFT-S-OFDM, and SS-SC-FDMA in detection reliability by selecting proper parameters as shown in Fig. \ref{Result_Case3_BER_MMSEPA_K1}. 

\begin{figure}[t]
\centering \centerline{
\includegraphics[width=0.5\textwidth,clip]{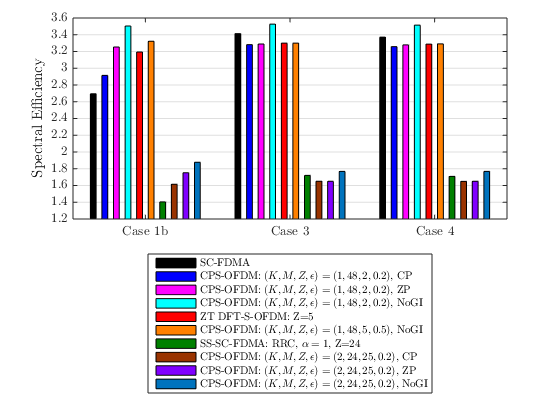}} 
\caption{Spectral efficiency comparisons of the waveforms claiming low PAPR in the interference-free single user case, the asynchronous multiuser case, and the mixed numerology multiuser case at $E_{\mathrm{b}}/N_{0}=25$ dB.}
\label{Result_SE_K1}
\end{figure} 

\subsection{Case 4: Mixed Numerology Multiuser Uplink Performance}
In Case 4, the performance of the target user in presence of two interfering users occupying adjacent subbands is studied.
The interfering users are synchronized with the target user, yet have different subcarrier spacing.
The detection performance of the target user applying different waveforms to the mixed numerologies in synchronism is given in this subsection. In Fig. \ref{Result_Case4_BER_MMSEPA_K2}, the superiority of CPS-OFDM over the other low OSBE waveforms can be found. Particularly, CPS-OFDM with ZP or NoGI exhibits robustness to the inter-numerology interference. In Fig. \ref{Result_Case4_BER_MMSEPA_K1}, it is obvious that CPS-OFDM can offer better detection reliability compared to the other low PAPR waveforms. Moreover, similar to the results in Fig. \ref{Result_Case3_BER_MMSEPA_K1} for Case 3, CPS-OFDM with the setting of $(K,M,Z,\epsilon)=(2,S/2,M+1,0.2)$ can obtain excellent BER performance mainly because of its extremely low OSBE as illustrated in Fig. \ref{Result_Case1b_PSDPA_K1}.

\subsection{Spectral Efficiency Analysis}
The spectral efficiency of applying different waveforms to Case 1b, Case 3, and Case 4 are summarized in this subsection. Let $T$ be the transmission time interval (TTI) which is $1$ ms. The spectral efficiency is defined as \cite{3GPPTR38802}
\begin{eqnarray}
\mathrm{SE} = \frac{\chi}{T(\mathrm{BW}_{\mathrm{UE}}+\Delta)},
\end{eqnarray}
where $\chi$ is the number of correctly received information bits by the target user per TTI, $\mathrm{BW}_{\mathrm{UE}}$ is the UE bandwidth, and $\Delta$ is the guard band size of the target user. According to \cite{Huawei166093_3GPPTDOC}, $\chi$ can be expressed in more detail as $\chi = N_{\mathrm{bit}}DN_{\mathrm{block}}(1-\mathrm{BER})$, where $N_{\mathrm{bit}}$ is the number of bits per QAM data symbol ($N_{\mathrm{bit}}=4$ for 16QAM), $D$ is the number of data symbols per block transmission, $N_{\mathrm{block}}$ is the number of transmitted blocks per TTI ($N_{\mathrm{block}}=14$ for CP and ZP, $N_{\mathrm{block}}=15$ for NoGI), and $\mathrm{BER}$ can be obtained from the results shown in Fig. \ref{Result_Case1b_BER_MMSEPA_K2} and Figs. \ref{Result_Case1b_BER_MMSEPA_K1}-\ref{Result_Case4_BER_MMSEPA_K1} at $E_{\mathrm{b}}/N_{0}=25$ dB of our interest. In Case 1b, $\Delta$ is chosen so that the OSBEP per $30$ kHz measurement bandwidth can be lower than the SEM requirement of $-18$ dBm. 
In Case 3 and Case 4, $\Delta$ is fixed to be $60$ kHz. As revealed in Fig. \ref{Result_SE_K2} and Fig. \ref{Result_SE_K1}, CPS-OFDM with NoGI achieves the highest spectral efficiency even though there exists the impairment of IBI at the receiver.

\section{Conclusion}
\label{Sec:Conclusion}
A new waveform called circularly pulse-shaped OFDM (CPS-OFDM), along with its optimal prototype vector design, is proposed for 5G New Radio (NR). CPS-OFDM possesses the advantages of both low out-of-subband emission (OSBE) and low peak-to-average power ratio (PAPR), which are two critical physical layer signal requirements of 5G air-interface. Thus, the spectral regrowth and the signal distortion caused by power amplifier (PA) nonlinearity can be substantially alleviated in contrary to OFDMA and other waveforms claiming low OSBE (e.g., WOLA-OFDM, UF-OFDM, f-OFDM, and OP-OFDM). Benefited from the design flexibility, CPS-OFDM can further reduce the PAPR so as to ensure better PA efficiency as compared to SC-FDMA and other waveforms claiming low PAPR (e.g., ZT DFT-S-OFDM and SS-SC-FDMA) by allowing little noise enhancement penalty (NEP). In addition, the proposed CPS precoder is able to be efficiently realized with linearithmic-order complexity through the characteristic matrix domain implementation method, which also identifies the condition of the CPS precoding matrix being unitary. Different from GFDM, CPS-OFDM endows every user in the system with flexibility to determine its own circular pulse shaping and orthogonal multiple access support. The optimal prototype shaping vector built in the CPS precoder is obtained by the proposed optimization procedure based on majorization-minimization (MM) and convex-iteration (CI) iteration processes. Simulation results demonstrate the performance gains in detection reliability and spectral efficiency of applying the proposed schemes to the practical sub-6-GHz uplink cases specified by 3GPP including single user interference-free communication, multiuser asynchronous transmissions, and multiuser mixed numerologies. According to these results, the superiority of CPS-OFDM over other waveform candidates is quite evident. We therefore consider the proposed CPS-OFDM to be one of the most promising technologies in 5G NR and beyond.

In the future, there are some remaining issues related to the use of CPS-OFDM and its optimization procedure for further study. The criteria to decide the parameters $K$, $M$, $\mathcal{K}$, $\mathcal{M}$, $\beta$, $\epsilon$, and $w$ adapting to various application scenarios are of interest. The downlink usage relying on multiple user-specific CPS precoders at the transmitter is to be investigated. Although the optimized prototype shaping vector obtained from the proposed Algorithm 1 can be offline computed and saved into a lookup table, one may accelerate the convergence rate of the MM process by finding a more suitable surrogate function alternative to (\ref{UBofObjFun}). A proof to the existence of the optimal rank-one solution of Problem (\ref{SDPProblem}) is anticipated. It is worthy to analyze the signal-to-interference ratio (SIR) of CPS-OFDM subject to multiuser interference arising from spectral regrowth, synchronization mismatch, and different subcarrier spacing for its advanced receiver development. Moreover, designing dedicated pilot sequences for CPS-OFDM channel estimation might be needed. Integrating the proposed CPS-OFDM system with MIMO technologies in an efficient way is strongly desirable and also essential to 5G air-interface.




\end{document}